\documentclass[11pt,a4paper]{article} 
\usepackage[utf8]{inputenc}

\usepackage[numbers]{natbib}
\usepackage{lettrine}
\usepackage{amsfonts}
\usepackage{amsmath} 
\usepackage{amssymb}  
\usepackage{amsthm}
\usepackage{dsfont}
\usepackage{epsfig} 
\usepackage{graphics} 
\usepackage{graphicx}
\usepackage{imakeidx}
\usepackage{mathptmx} 
\usepackage{mathtools}
\usepackage{natbib}
\usepackage{siunitx}
\usepackage{subcaption}
\usepackage{times} 
\usepackage{verbatim}
\usepackage{algpseudocode}
\usepackage{tikz}
\usepackage{hyperref}
\usetikzlibrary{arrows,automata,positioning,intersections}
\usepackage[pdf]{graphviz} 
\usepackage{verbatim}

\usepackage[english]{babel}

\newtheorem{thrm}{Theorem}[section]

\newtheorem{prpstn}[thrm]{Proposition}

\newtheorem{lgrthm}[thrm]{Algorithm}

\newtheorem{dfntn}[thrm]{Definition}

\newtheorem{xmpl}[thrm]{Example}
\newtheorem{prblm}[thrm]{Problem}
\newtheorem{rmrk}[thrm]{Remark}

\newcommand{\argmin}{\operatorname{argmin}}
\newcommand{\sgn}{\operatorname{sign}}

\newcommand{\Qc}{\mathcal{Q}}
\newcommand{\Wc}{\mathcal{W}}

\newcommand{\BBb}{\mathbb{B}}
\newcommand{\Ebb}{\mathbb{E}}
\newcommand{\Gbb}{\mathbb{G}}
\newcommand{\Vbb}{\mathbb{V}}

\newcommand{\Prec}{\operatorname{Prec}}
\newcommand{\Real}{\mathbb{R}}
\newcommand{\Suff}{\operatorname{Suff}}

\title{\LARGE \bf The Bounded Acceleration Shortest Path problem:
  complexity and solution algorithms}

\author{Stefano Ardizzoni, Luca Consolini, Mattia Laurini
and Marco Locatelli
\thanks{All authors are with the Dipartimento di Ingegneria
e Architettura, Universit\`a degli Studi di Parma,
Parco Area delle Scienze 181/A, 43124 Parma, Italy. E-mails:
{\tt\footnotesize \{stefano.ardizzoni, luca.consolini, mattia.laurini,
marco.locatelli\}@unipr.it}}%
}
\date{}

\begin{document}
 
\maketitle

\begin{abstract}
The purpose of this work is to introduce and characterize the Bounded Acceleration
Shortest Path
(BASP) problem, a generalization of the Shortest Path (SP)
problem. This problem is associated to a graph: the nodes represent
positions of a mobile vehicle and
the arcs are associated to pre-assigned geometric paths that connect these positions.
BASP consists in finding the minimum-time path between
two nodes. Differently from SP, we require that
the vehicle satisfy bounds on maximum and minimum acceleration and
speed, that depend on the vehicle position on the currently traveled
arc. We prove that BASP is NP-hard and define solution
algorithm that achieves polynomial time-complexity under some
additional hypotheses on problem data.

\end{abstract}

%

\section{Introduction}

\begin{figure}[!tbh]
\centering
\includegraphics[width=0.6\columnwidth]{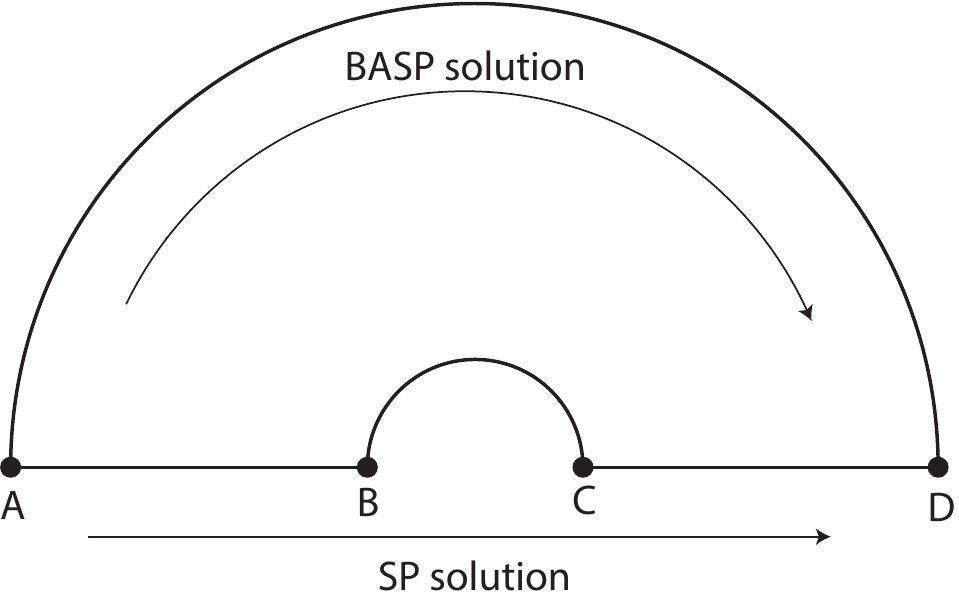}
\caption{Comparison of BASP and SP solutions.}
\label{fig:percorso_esempio}
\end{figure}

\begin{figure}[!tbh]
\centering
\includegraphics[width=0.7\columnwidth]{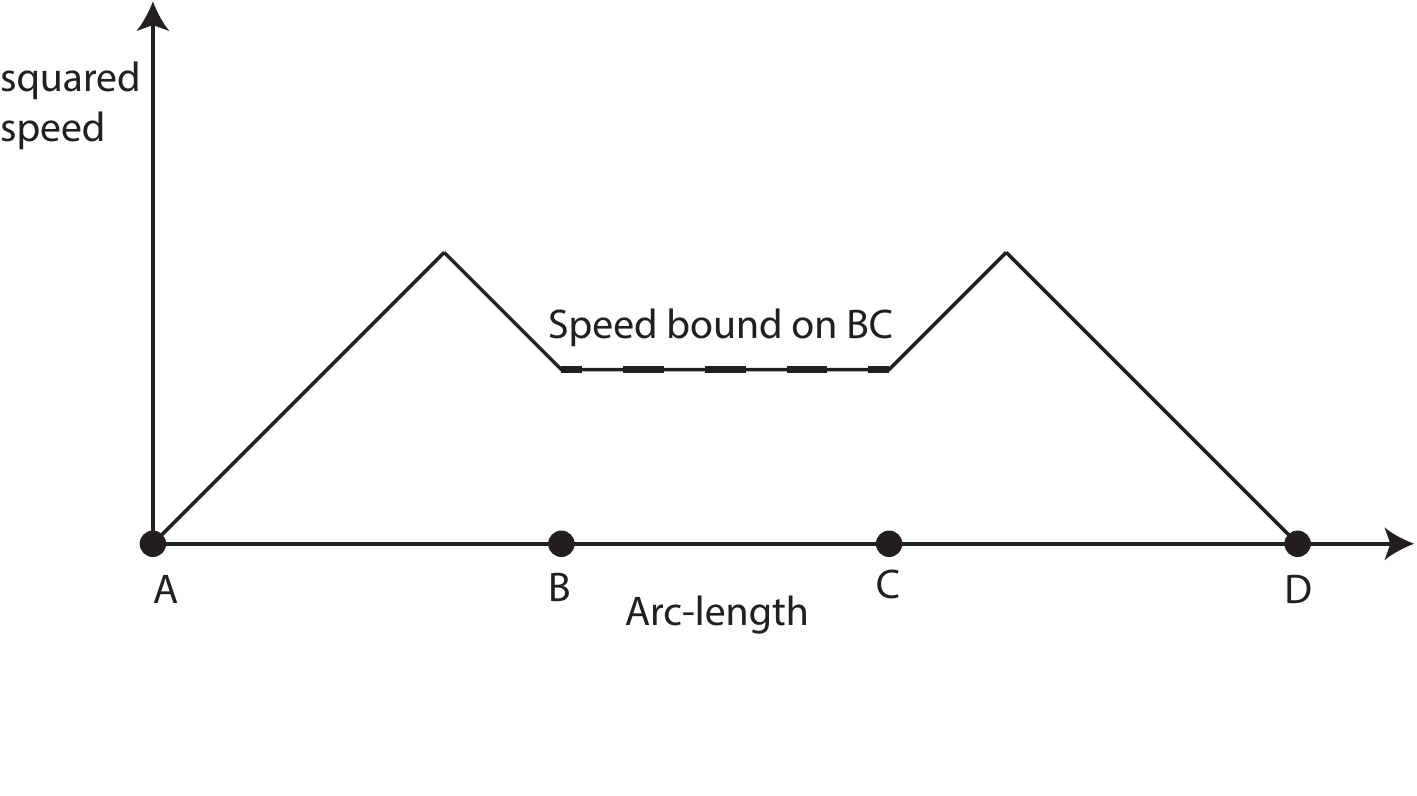}
\caption{Optimal speed profile on $ABCD$.}
\label{fig:veloc_ABCD}
\end{figure}

\begin{figure}[!tbh]
\centering
\includegraphics[width=0.7\columnwidth]{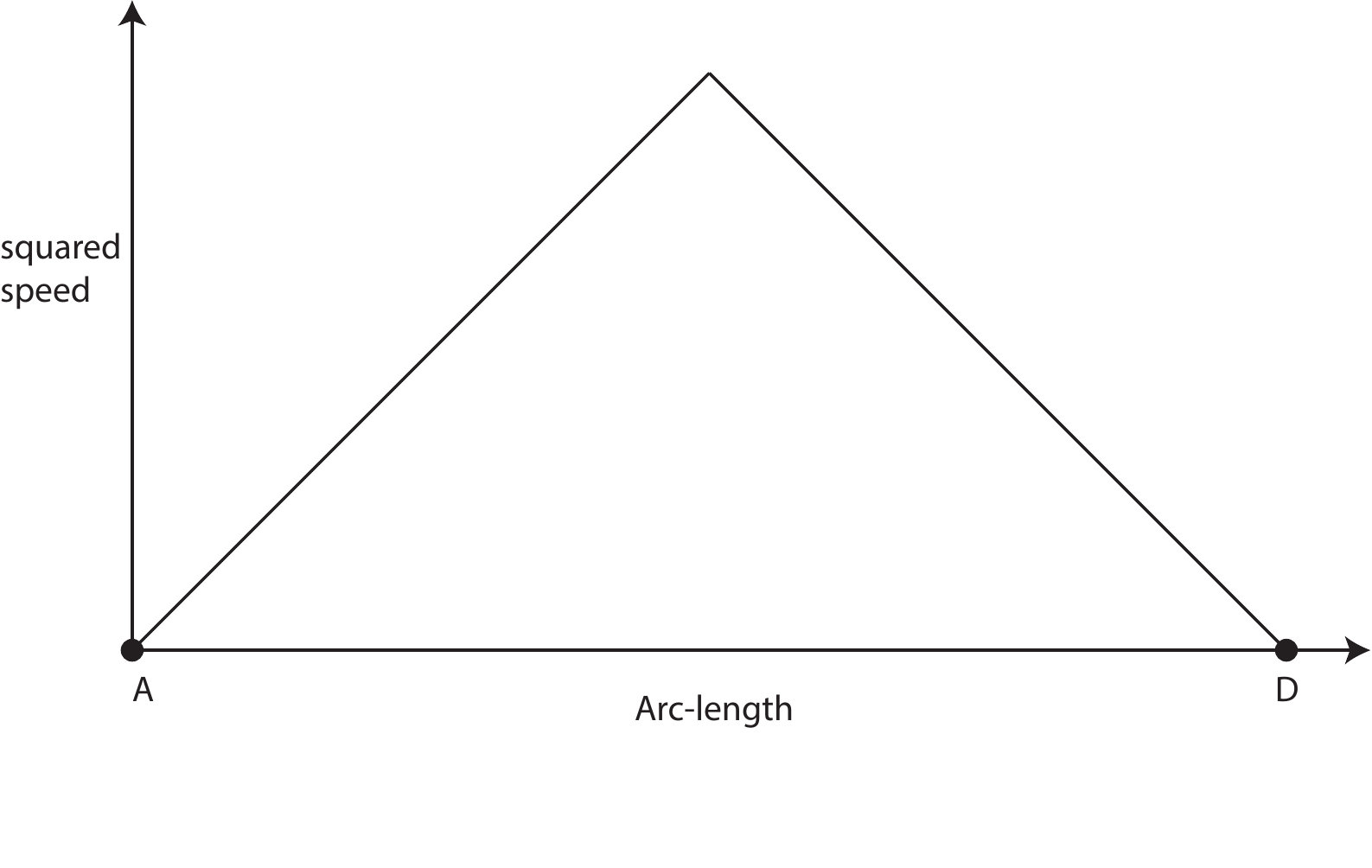}
\caption{Optimal speed profile on $AD$.}
\label{fig:veloc_AD}
\end{figure}

The purpose of this work is to introduce and characterize the Bounded Acceleration
Shortest Path
(BASP) problem, a generalization of the Shortest Path (SP)
problem. We consider a graph associated to a path and speed planning problem
for a mobile vehicle. The graph nodes represent vehicle positions and
the arcs are associated to pre-assigned geometric paths that connect these positions.
BASP consists in finding the minimum-time path between
two nodes. Differently from SP, BASP requires that
the vehicle satisfy bounds on maximum and minimum acceleration and
speed, that depend on the vehicle position on the currently traveled
arc. Figure~\ref{fig:percorso_esempio} presents a simple scenario that
allows to illustrate BASP and
its difference with SP. This figure shows some fixed paths connecting
positions $A,B,C,D$. The vehicle starts from $A$ with zero speed and must reach $D$
with zero speed. The solution of the SP problem corresponds to path $ABCD$,
which is the one of shortest length. 
BASP consists in finding the shortest-time path under acceleration and
speed constraints. In this case, we assume
that the vehicle acceleration and deceleration are bounded by a common
constant and that its speed is bounded only on arc $BC$. For
instance, this may be due to the fact that $BC$ is an arc of a
circle of small radius and the vehicle speed on $BC$ has to be limited
in order to avoid excessive lateral acceleration, which may cause
slideslip. If the bound on acceleration and deceleration is sufficiently high, the solution of BASP
corresponds to path $AD$. Indeed, even if this second path is
longer, it can be travelled with a greater mean speed due to the
absence of speed bounds. To clarify this fact, see
Figures~\ref{fig:veloc_ABCD} and~\ref{fig:veloc_AD}. Figure~\ref{fig:veloc_ABCD} represents the
fastest speed profile on $ABCD$. The $x$-axis corresponds to the
arc-length position on path $ABCD$ and the $y$-axis represents the
squared speed. In this
representation, arc-length
intervals of constant acceleration or deceleration correspond to straight lines.
Note that speed has to be reduced
before entering into arc $BC$ in order to respect the speed bound on
$BC$. Figure~\ref{fig:veloc_AD} represents the fastest speed profile
on $AD$. Due to the absence of speed bounds, the vehicle accelerates
till the midpoint of
the path and then decelerates to the end node $D$. Even if path $AD$
is longer than $ABCD$, it can be travelled with a shorter time.
In Section~\ref{sec_assigned_path}, we will justify the structure of
the optimal speed profiles reported in Figures~\ref{fig:veloc_ABCD} and~\ref{fig:veloc_AD}.

BASP is a generalization of SP. Indeed, if we remove the maximum and
minimum acceleration bounds, BASP reduces to SP, in which the cost
of each arc is the time needed to travel the path
associated to the arc at maximum speed.
In the general case, BASP is more complex than SP. Indeed, 
in Proposition~\ref{prop_NP_hard}, we will show that BASP is NP-hard.
However, if we make some additional assumptions on problem parameters,
we obtain a subclass of BASP, that we call $k$-BASP, that can be
solved with polynomial time-complexity.
Roughly speaking, a BASP instance belongs to $k$-BASP if the problem
data are such that no more than $k-2$ arcs can be travelled with a speed
profile starting from zero speed and of maximum acceleration, then followed
by one of maximum deceleration and ending with zero speed, without violating
the maximum speed constraint.
In Section~\ref{sec_kBASP}, we will define $k$-BASP more precisely and we
will present a simple upper bound on constant $k$.
In Proposition~\ref{prop_complexity}, we will show that $k$-BASP can be solved
by Dijkstra's algorithm with polynomial time complexity with respect to the graph
size (its number of nodes and edges), provided that $k$ is fixed.
We will also present Algorithm~\ref{alg_enh_ad}, which is able to adaptively find
constant $k$. 

{\bf Statement of contribution.}
To our knowledge, BASP has not been explicitly considered in literature,
so that the main results presented here are new.
In particular, we believe that the most relevant contributions are:
\begin{itemize}
\item Proposition~\ref{prop_NP_hard}, that shows that BASP is NP-hard.
\item Proposition~\ref{prop_complexity}, that shows that $k$-BASP can be solved
with a polynomial time-complexity, provided that $k$ is fixed.
\item The adaptive Algorithm~\ref{alg_enh_ad}, that is able to efficiently solve a
large set of BASP instances.
\end{itemize}

\subsection{Problem motivation}
One relevant application of this work is the optimization of automated
guided vehicles (AGVs) motions in automated warehouses.
Automated warehouses are rapidly spreading in manufacturing and
logistics because of their speed, flexibility, and reliability. 
In order to ensure the smooth functioning and to increase the overall
efficiency of the system, such fleets of AGVs need be coordinated at
different levels of control: task allocation, localization, path planning,
motion planning and vehicle management (see, for instance,~\cite{RYCK2020},
for a more in depth discussion).

\begin{figure}[!tbh]
\centering
\includegraphics[width=\columnwidth]{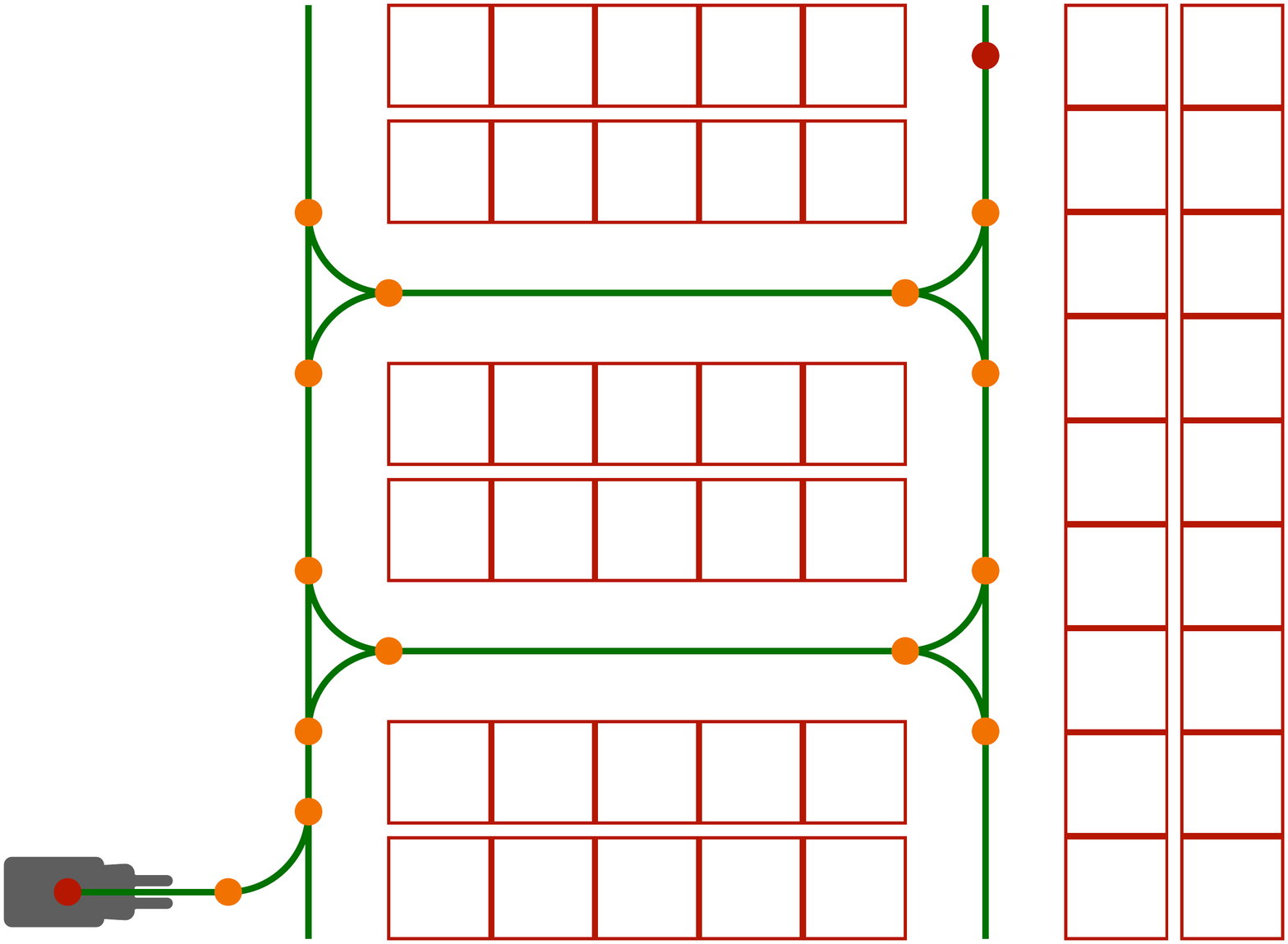}
\caption{An example scenario.}
\label{fig:warehouse}
\end{figure}

In automated warehouses, AGVs are commonly moved between fixed operating points.
These points may be associated to shelves locations, where
packages are stored or retrieved, to the end of production lines, where
the AGV picks up a final product, and to additional intermediate locations,
used for routing. 
Between these operating points, the vehicle follows preassigned
connecting paths (see Figure~\ref{fig:warehouse}).
The vehicle motion must satisfy constraints on maximum
speed and maximum tangential and transversal accelerations, that
depend on the vehicle position on the path.

The algorithms developed in this paper allow to find the
time-optimal path for a single AGV that travels between operating
points, taking into account path-dependent bounds on maximum
acceleration, deceleration and speed.

\subsection{Related works}
As said, to our knowledge, BASP has not been explicitly addressed in literature.
However, various works address related path planning problems for AGVs.
In those scenarios in which AGVs are allowed to move freely within their
environment and no predetermined circuits are available, one need to
employ environmental representations such as cell decomposition methods
(\cite{AHMAD2015}) or trajectory maps (\cite{SEIF2015}).
In particular, among cell decomposition methods,~\cite{COWLAGI2009} presents
an algorithm based on a modification of Dijkstra's algorithm in which edge weights
depend on previously visited edges.
Note that our work shares some similarities with~\cite{COWLAGI2009} in regards
of the idea of the history-dependent edges weight and in the way the extended
graph associated to the addressed problem is defined.
However,~\cite{COWLAGI2009} focuses on a different problem. In fact,
it introduces a cell decomposition method whose
goal is to obtain a feasible path taking into account the vehicle maximum curvature
radius. Instead, our work focuses on selecting the optimal path among
a set of already feasible paths while obtaining the optimal speed profile as well.
Moreover, the algorithm introduce in~\cite{COWLAGI2009} operates introducing
a set of labels which can potentially be very expensive in terms of memory usage
and the history parameter is given in input and is not adaptively computed, losing
the guarantee of optimality.

In many industrial scenarios, AGVs move along predetermined circuits.
The representation of such paths is usually graph-based.
The problem of finding the optimal path connecting two positions within a
facility turns, then, into the problem of finding the shortest path connecting
a pair of nodes in a graph.
There are various graph searching algorithms that are used to this end such
as A$^*$, Lifelong Planning A$^*$ (\cite{KOENIG2004}),
D$^*$ (\cite{FERGUSON2007}) and D$^*$ Lite algorithms.
Among these, the most widely used (\cite{KIM2018}) are
A$^*$ and D$^*$ Lite algorithms.

A$^*$ algorithm (\cite{NILSSON1971}) is a heuristic method
that allows to compute the optimal path (if it exists) (\cite{GELPERIN1977})
by exploring the graph beginning from the starting node along the most promising
directions according to a heuristic function that estimates the cost from the current
position to the target node.

\section{Notation}
A directed graph is a pair $\Gbb=(\Vbb,\Ebb)$ where $\Vbb$ is a set of
nodes
and $\Ebb \subset \{(x,y) \in \Vbb^2 \mid x \neq y\}$ is a set of directed arcs.
A path $p$ on $\Gbb$ is a sequence of adjacent vertices of $\Ebb$.
That is, $p= \sigma_{1}\cdots \sigma_{m}$, where, for $i \in \{1,\ldots,m-1\}$,
$(\sigma_i,\sigma_{i+1}) \in \Ebb$.
We denote by $P(\Gbb)$ the set of all paths of $\Gbb$.
An alphabet $\Sigma = \{\sigma_1,\ldots,\sigma_n\}$ is a set whose elements
are called symbols.
A word is any finite sequence of symbols.
We denote the set of all words over $\Sigma$ by $\Sigma^*$, that also contains
the empty word $\epsilon$, while $\Sigma_i$ represents the set of all words of
length up to $i \in \mathbb{N}$, that is, words composed of up to $i$ symbols,
including the empty word $\epsilon$.
Given a word $w \in \Sigma^*$, we denote its length by $|w|$.
Given a directed graph $\Gbb = (\Vbb, \Ebb)$, we can think of $\Vbb$ as
an alphabet.
In this way, any path $p \in P(\Gbb)$ is a word in $\Vbb^*$.
Given $s,t \in \Sigma^*$, the word obtained by writing $t$ after $s$ is
called the concatenation of $s$ and $t$ and is denoted by $st \in
\Sigma^*$. We also say that $t$ is a suffix of $st$ and that $s$ is a
prefix of $st$.
For $r \in \Vbb^*$, we denote by $\vec{r}$ the rightmost symbol of $r$.
In the following, it will be convenient to represent paths of $\Gbb$
as strings composed of symbols in $\Vbb$.
This will allow us to use the concatenation operation on paths and to
use prefixes and suffixes to represent portions of paths.

For $x \in \Real$, we denote the ceiling of $x$ by $\lceil x \rceil =
\min \{i \in \mathbb{Z} \mid i \geq x\}$.
For $a,b \in \Real$, we set $a \wedge b = \min\{a,b\}$ and $a \vee b =
\max\{a,b\}$, as the minimum and maximum operations, respectively. 
Further, $\Real^+$ denotes the set of nonnegative real numbers.
 
Finally, given an interval $I \subseteq \Real$, let us recall that $W^{1,\infty}(I)$
is the Sobolev space of functions in $L^{\infty}(I)$ with weak derivative
of order one with finite $L^{\infty}$-norm.
For $f, g \in W^{1,\infty}(I)$, we denote with $f \wedge g$ and $f \vee g$
the point-wise minimum and maximum of $f$ and $g$, respectively.

\section{Problem formulation}
We first present the speed planning problem on an assigned path,
following
our previous work~\cite{CONSOLINI2020}. Then, we introduce the BASP
problem, that considers both speed planning and path selection.
\subsection{Speed planning along an assigned path}
\label{sec_assigned_path}
Let $\gamma: \left[ 0, \lambda_f \right] \rightarrow \Real^2$ be a
$C^2$ function such that $(\forall \lambda \in \left[ 0, \lambda_f \right])\
\left\| \gamma^\prime(\lambda) \right\| = 1$.
The image set $\gamma\left(\left[0,\lambda_f\right]\right)$ represents
the path followed by a vehicle, $\gamma(0)$ the initial configuration and
$\gamma(\lambda_f)$ the final one.
Function $\gamma$ is an arc-length parameterization of the path.
We want to compute the speed-law that minimizes the overall travel time
while satisfying some kinematic and dynamic requirements.
To this end, let $\xi: \left[ 0, t_f \right] \rightarrow \left[ 0, \lambda_f \right]$
be a differentiable monotone increasing function that represents the vehicle
arc-length coordinate along the path as a function of time and let
$v: \left[ 0, \lambda_f \right] \rightarrow [ 0, +\infty)$ be such that,
$(\forall t \in \left[0,t_f\right])\ \dot{\xi}(t) = v(\xi(t)).$
In this way, $v(\lambda)$ is the vehicle speed at position $\lambda$.
The position of the vehicle as a function of time is given by $x:
\left[ 0, t_f \right] \rightarrow \Real^2$, $x(t) = \gamma(\xi(t))$, speed and
acceleration are given by
\begin{gather*}
\dot{x}(t) = \gamma^\prime(\xi(t))v(\xi(t)),\\
\ddot{x}(t) = a_L(t)\gamma^\prime(\xi(t)) + a_N(t)\gamma^{\prime\perp}(\xi(t)),
\end{gather*}
where $a_L(t) = v^\prime(\xi(t))v(\xi(t))$ and $a_N(t)(t)= \kappa (\xi(t))v(\xi(t))^2$
are the longitudinal and normal components of acceleration, respectively.
Here, $\kappa: \left[ 0, \lambda_f \right] \rightarrow \Real$ is the scalar curvature,
defined as $\kappa(\lambda) = \left\langle \gamma^{\prime\prime}(\lambda),
\gamma^{\prime}(\lambda)^\perp \right\rangle$, where $\langle\cdot,\cdot\rangle$
denotes the scalar product.

We require to travel distance $\lambda_f$ in minimum-time while
satisfying, for every $t \in [0,\xi^{-1} (\lambda_f)]$, $0 \leq v^-(\xi(t)) \leq
v(\xi(t)) \leq v^+(\xi(t))$, $|a_N(\xi(t))| \leq \beta(\xi(t))$, $\alpha^-(\xi(t))
\leq a_L(\xi(t)) \leq \alpha^+(\xi(t))$.
Here, functions $v^-,v^+,\alpha^-,\alpha^+,\beta$ are arc-length dependent
bounds on the vehicle speed and on its longitudinal and normal acceleration.
It is convenient to make the change of variables $w = v^2$
(see~\cite{Verscheure09}), so that our problem takes on the following form
\begin{subequations}
\label{eqn_problem_pr_w}
\begin{align}
\min_{w \in W^{1, \infty}\left([0,\lambda_f]\right)}
& \int\limits_0^{\lambda_f} w(\lambda)^{-\frac{1}{2}} d \lambda
& \label{obj_fun_pr_w} \\
& \mu^-(\lambda) < w(\lambda) \leq \mu^+(\lambda),
& \lambda \in [0,\lambda_f], \label{con_speed_pr_w} \\
& \alpha^-(\lambda) \leq w'(\lambda) \leq \alpha^+(\lambda),
& \lambda \in [0,\lambda_f], \label{con_at_pr_w}
\end{align}
\end{subequations}
where
\begin{equation}
\label{eqn_u}
\mu^+(\lambda) = v^+(\lambda)^2 \wedge \frac{\beta(\lambda)}{\kappa(\lambda)},
\qquad \mu^-(\lambda)= v^-(\lambda)^2
\end{equation}
represent the upper bound on $w$ (depending on speed bound $v^+$ and
curvature $\kappa$) and the lower bound on $w$, respectively.

We actually address the following problem,
which is slightly more general than~\eqref{eqn_problem_pr_w},
\begin{subequations}
\label{eqn_problem_pr_wg}
\begin{align}
\min_{w \in W^{1,\infty}\left([0,\lambda_f]\right)}
&\Psi(w)
& \label{obj_fun_pr_wg} \\
& \mu^-(\lambda) \leq w(\lambda) \leq  \mu^+(\lambda),& \lambda \in [0,\lambda_f],
\label{con_speed_pr_wg} \\
& \alpha^-(\lambda) \leq w'(\lambda) \leq \alpha^+(\lambda),
& \lambda \in [0,\lambda_f], \label{con_at_pr_wg}
\end{align}
\end{subequations}
where $\Psi: W^{1,\infty}\left(\left[0,\lambda_f\right]\right) \to \Real$
is order reversing (i.e., $(\forall x, y \in \left[0,\lambda_f\right])\
x \geq y \Rightarrow \Psi(x) \leq \Psi(y)$) and $\mu^-$, $\mu^+$, $\alpha^-$,
$\alpha^+ \in L^\infty\left(\left[0,\lambda_f\right]\right)$ are assigned functions
with $\mu^-, \alpha^+\geq 0$ and $\alpha^-\leq 0$.
Initial and final conditions on speed can be included in the
definition of functions $\mu^-$ and $\mu^+$. For instance, to set
initial condition $w(0)=w_0$, it is sufficient to define $\mu^+(0)=\mu^-(0)=w_0$.

Note that the objective function~\eqref{obj_fun_pr_w} is order reversing, so that
Problem~\eqref{eqn_problem_pr_w} has form~\eqref{eqn_problem_pr_wg}.

\subsection{Solution of Problem~\eqref{eqn_problem_pr_wg}}
    
We summarize the method presented in~\cite{CONSOLINI2020} and
begin with introducing a subset of $W^{1,\infty}([0,\lambda_f])$ as a
technical requirement.

\begin{dfntn}
\label{def:defQ}
Let $Q_{\alpha^-,\alpha^+}$ be the subset of $W^{1,\infty}([0,\lambda_f,])$ such
that $\mu \in Q$ if $\sgn\left(\mu^\prime - \alpha^+\right)$
and $\sgn\left(\mu^\prime - \alpha^-\right)$ are Riemann integrable
(i.e., in view of the boundedness of the $\sgn$ function, almost
everywhere continuous), where $\sgn: \Real \rightarrow \{-1, 0, 1\}$
is defined as\[
\sgn(x) =
\begin{cases}
1, & \mbox{if } x > 0 \\
0, & \mbox{if } x = 0 \\
-1, & \mbox{if } x < 0.
\end{cases}
\]
\end{dfntn}

Note that $\mu \in Q_{\alpha^-,\alpha^+}$ if functions $\mu^\prime - \alpha^+$ and
$\mu^\prime - \alpha^-$ change sign a finite number of times in
interval $[0,\lambda_f]$. In the following, we assume that $\mu^+ \in
Q_{\alpha^-,\alpha^+}$.

To solve Problem~\eqref{eqn_problem_pr_wg}, we define operators
$F, B, M : Q\rightarrow W^{1,\infty}([0,\lambda_f])$ where $Q$ is
defined as in Definition \ref{def:defQ} and, for $\mu \in Q$, $F(\mu)$
and $B(\mu)$ are given as follows
\begin{equation}
\label{F_def}
\begin{cases}
F(\mu)'(\lambda) =
\begin{cases}
\alpha^+(\lambda) \wedge \mu'(\lambda) & \!\text{if } F(\mu)(\lambda) \geq \mu (\lambda) \\
\alpha^+(\lambda) 				& \!\text{if } F(\mu)(\lambda) < \mu(\lambda)
\end{cases}\\
F(\mu)(0) = \mu(0),
\end{cases}
\end{equation}
\begin{equation}
\label{B_def}
\begin{cases}
B(\mu)'(\lambda) =
\begin{cases}
\alpha^-(\lambda) \wedge \mu'(\lambda) & \text{if } B(\mu)(\lambda) \geq \mu (\lambda) \\
\alpha^-(\lambda) 				& \text{if } B(\mu)(\lambda) < \mu(\lambda)
\end{cases}\\
B(\mu)(\lambda_f) =  \mu(\lambda_f).
\end{cases}
\end{equation}

\noindent
Finally, for $\mu \in Q$, operator $M$ is defined as
\begin{equation}\label{M_def}
M(\mu)=F(\mu)\wedge B(\mu).
\end{equation}
The solution of Problem~\eqref{eqn_problem_pr_wg} is given by
$w=M(\mu^+)$ (see~\cite{CONSOLINI2020}).
We call $F,B,M$ the forward operator, the backward operator and the meet
operator, respectively.
Roughly speaking, given a maximum squared speed profile $\mu^+\in Q$,
starting from $\mu^+(0)$  and up to $\mu^+(\lambda_f)$, $F(\mu^+)$ grows
with the maximum allowed acceleration $\alpha^+$ while staying below
$\mu^+$, and, if it touches $\mu^+$, coincides with it until $\mu^+$ grows
with an acceleration higher  than $\alpha^+$, in which case $F(\mu^+)$
behaves again as previously explained.
Analogously, operator $B$ acts in the same way as $F$ but backwards and
with $-\alpha^-$ as maximum acceleration.
Finally, meet operator $M$ is the point-wise minimum between forward
operator $F$ and backward operator $B$.
Moreover, Problem~\eqref{eqn_problem_pr_wg} is feasible if and only if
$\mu^- \leq w$.

\begin{figure}[!h]
\centering
\includegraphics[width=.65\columnwidth, trim=67 2 114 24, clip=true]{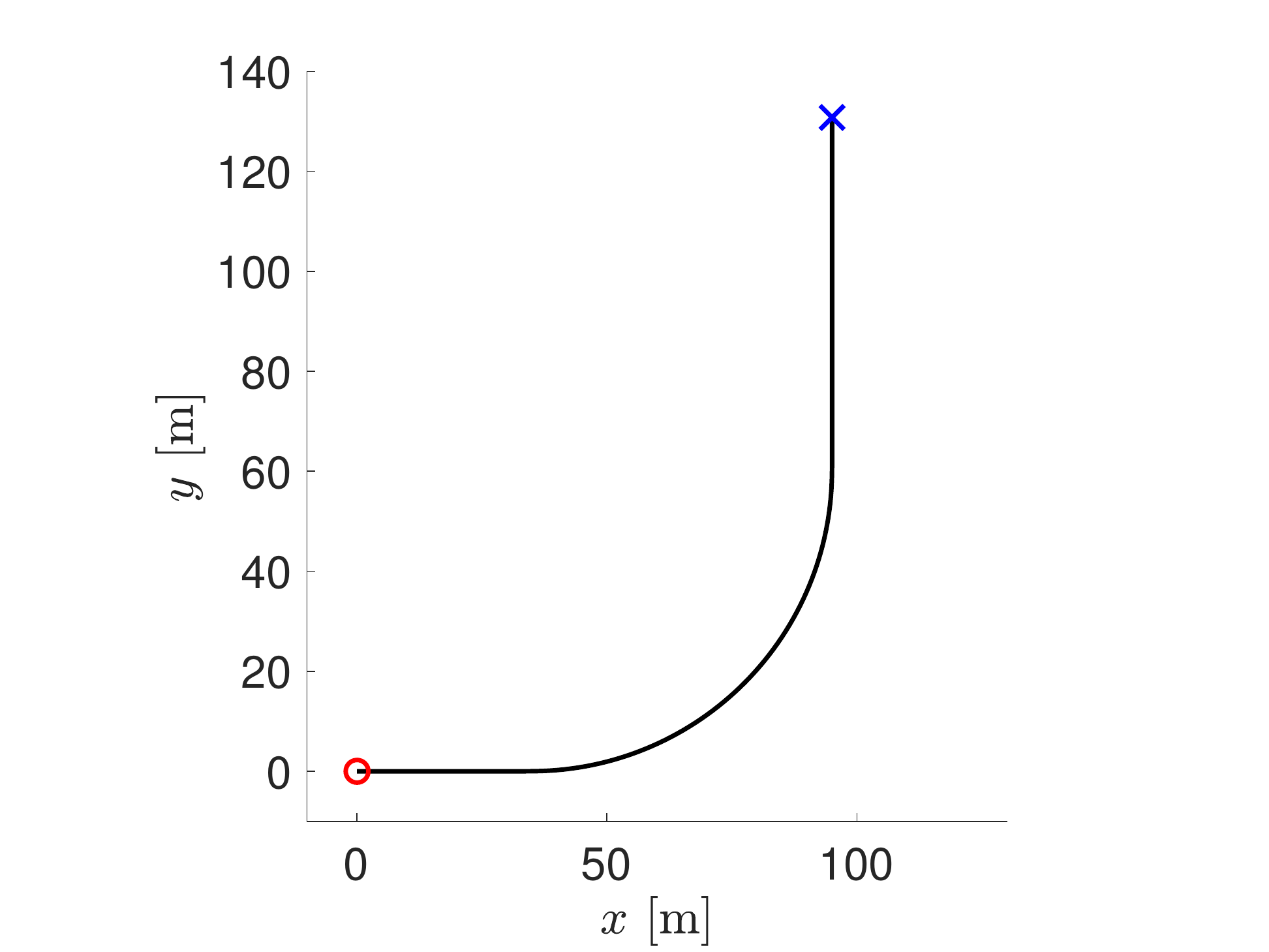}
\caption{The black line represents the path, while the red circle and the
black cross represent the starting point and the end point, respectively.}
\label{fig:path}
\end{figure}

\begin{figure}[!h]
\centering
\includegraphics[width=.9\columnwidth, trim=0 5 0 25, clip=true]{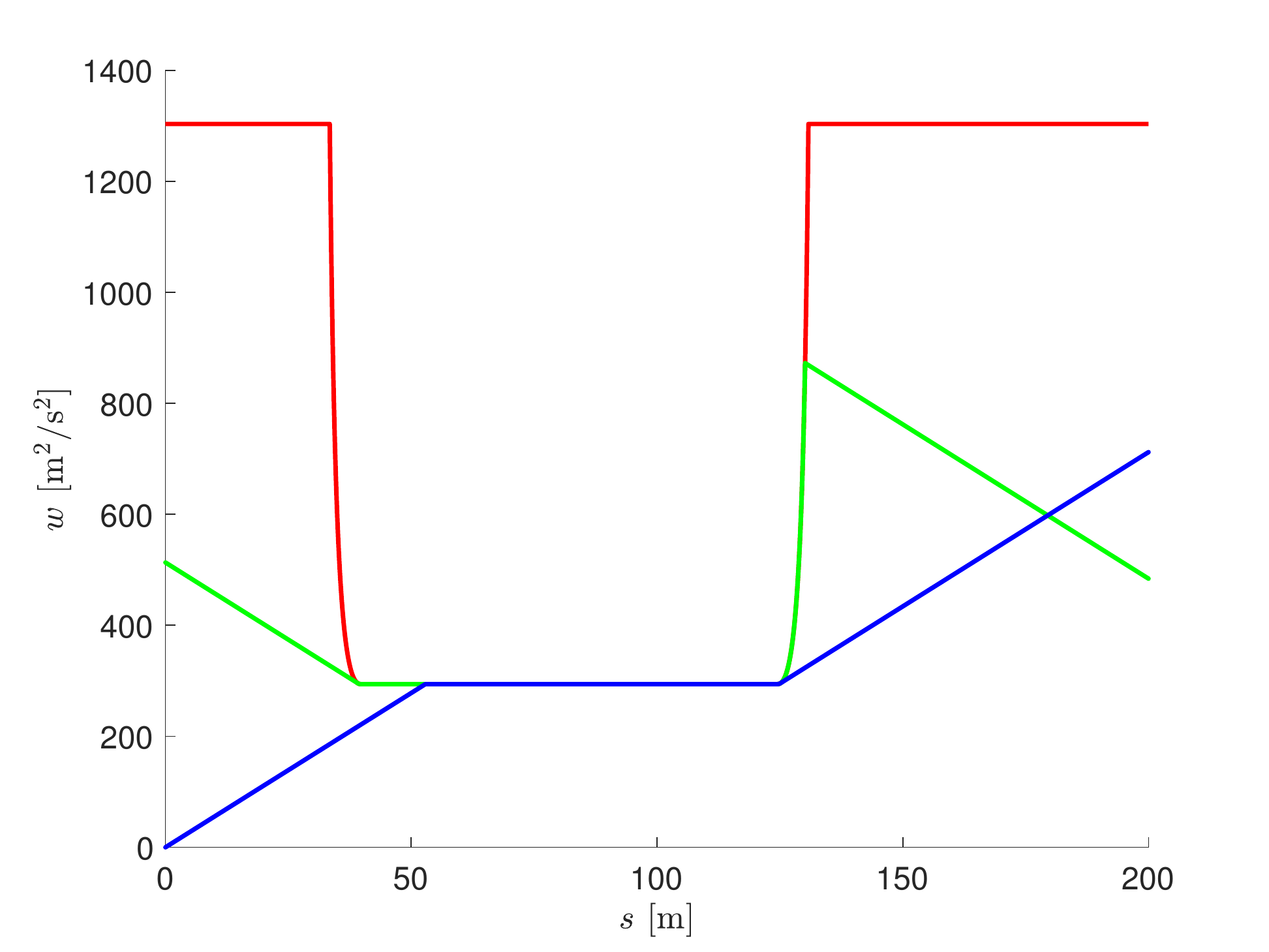}
\caption{The red line represents $\mu^+$ defined in~\eqref{eqn_u},
the blue one represents $F(\mu^+)$, whilst the green one represents
$B(\mu^+)$.}
\label{fig:operators_f_b}
\end{figure}

\begin{figure}[!h]
\centering
\includegraphics[width=.9\columnwidth, trim=0 5 0 25, clip=true]{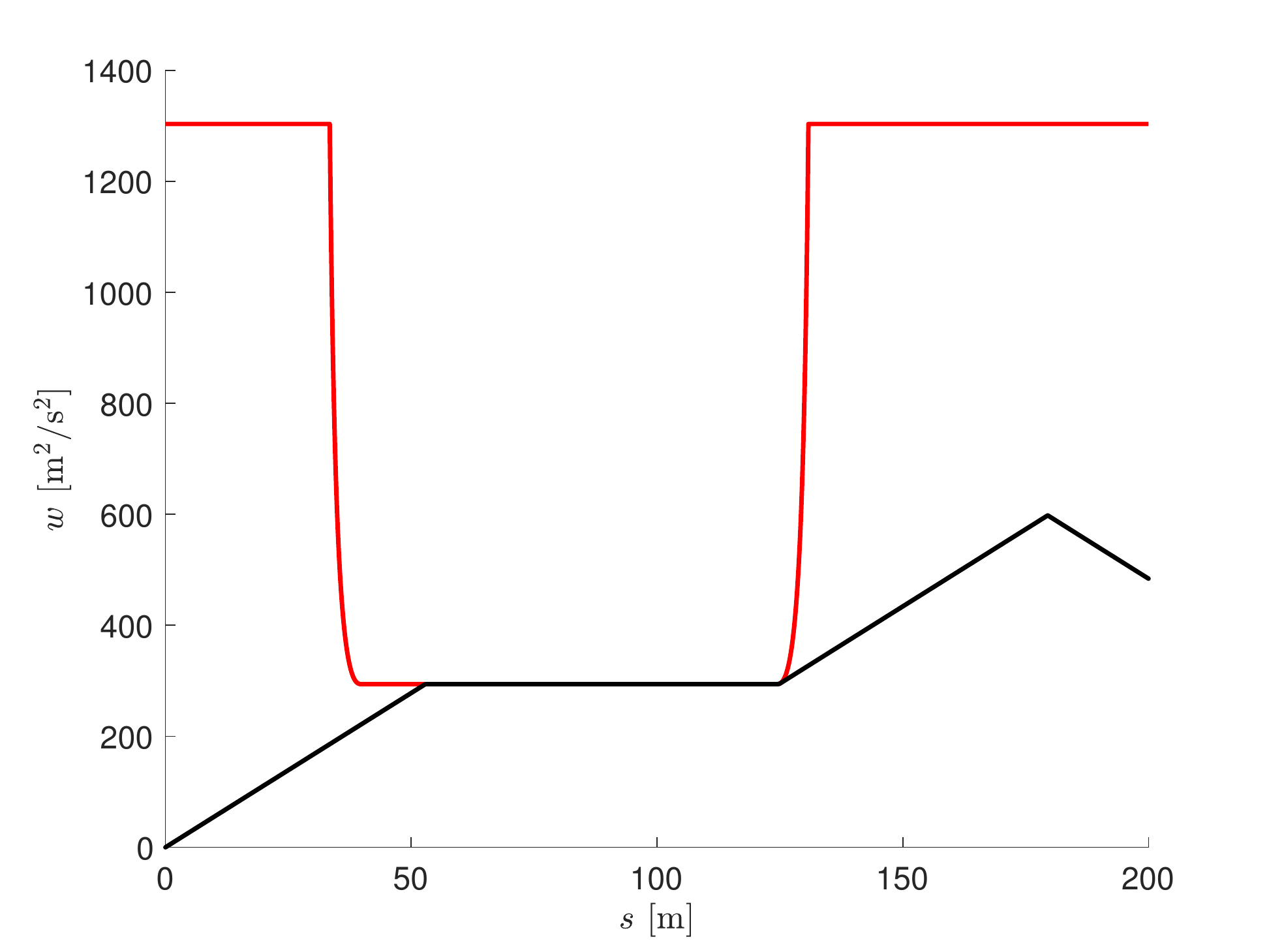}
\caption{Example 1: The red line represents $\mu^+$ defined in~\eqref{eqn_u},
whilst the black one represents optimal solution $w = M(\mu^+)$.}
\label{fig:operator_m}
\end{figure}

In order to further clarify the meaning of these operators, we will consider a
simple example.
Let us examine the path shown in Figure~\ref{fig:path}, which represents
a path whose total length is $200$ \si{\metre}.
The speed bounds $v^{+}$ and $v^{-}$ in~\eqref{eqn_u} are set as follows:
$v^{+}(0) = v^{-}(0) = 0$ \si{\metre\per \second}, $v^{+}(200) = v^{-}(200) = 22$
\si{\metre\per \second}, whilst, for each $\lambda \in (0 , 200)$, $v^{-}(\lambda) = 0$
\si{\metre\per \second} and $v^{+}(\lambda) = 36.1$ \si{\metre\per \second}.
The longitudinal acceleration limits are $\alpha^-= -2.78$ \si{\metre\per \square\second}
and $\alpha^+ = 2.78$ \si{\metre\per \square\second}, and the maximal normal
acceleration is $\beta = 4.9$ \si{\metre\per \square\second}.
Figure~\ref{fig:operators_f_b} shows the upper-bound function
$\mu^+$ obtained by~\eqref{eqn_u}, with $\mu^+(0)=0$, to impose zero
initial speed, and the corresponding functions
$F(\mu^+)$ and $B(\mu^+)$ computed as the solution of~\eqref{F_def} and~\eqref{B_def}, respectively. Figure~\ref{fig:operators_f_b} shows
$F(\mu^+)$ and $B(\mu^+)$, while Figure~\ref{fig:operator_m} shows the optimal solution $w = M(\mu^+)$
obtained by \eqref{M_def}.
In this example, the initial speed is zero, then the profile grows to the
upper bound $\mu^+$; next, it follows it in order to respect the
maximum speed constraint due to the lateral acceleration on the curve.
After that, at the end of the path part of higher curvature, it grows again and reaches
a second local maximum speed after which it decreases in order to meet the
final speed requirement $v^+(200)$.
Note that we can compute an approximated solution of $w=M(\mu^+)$ by
using a finite difference approximation of equations~\eqref{F_def}
and~\eqref{B_def}. As shown in~\cite{CONSOLINI201750}, this can be
done with an algorithm that has linear time-complexity with respect to
the number of discretization points.
Further, note that if functions $\mu^+$, $\alpha^-$, $\alpha^+$ are
piecewise-constant, then $w$ is piecewise linear (as in the simple
examples of Figures~\ref{fig:veloc_ABCD} and~\ref{fig:veloc_AD}) and
can be directly computed without actually integrating differential
equations~\eqref{F_def} and~\eqref{B_def}. 

\subsection{Bounded Acceleration Shortest Path Problem}

\begin{figure}[!h]
\centering
\includegraphics[width=.8\columnwidth, trim=0 30 0 13, clip=true]{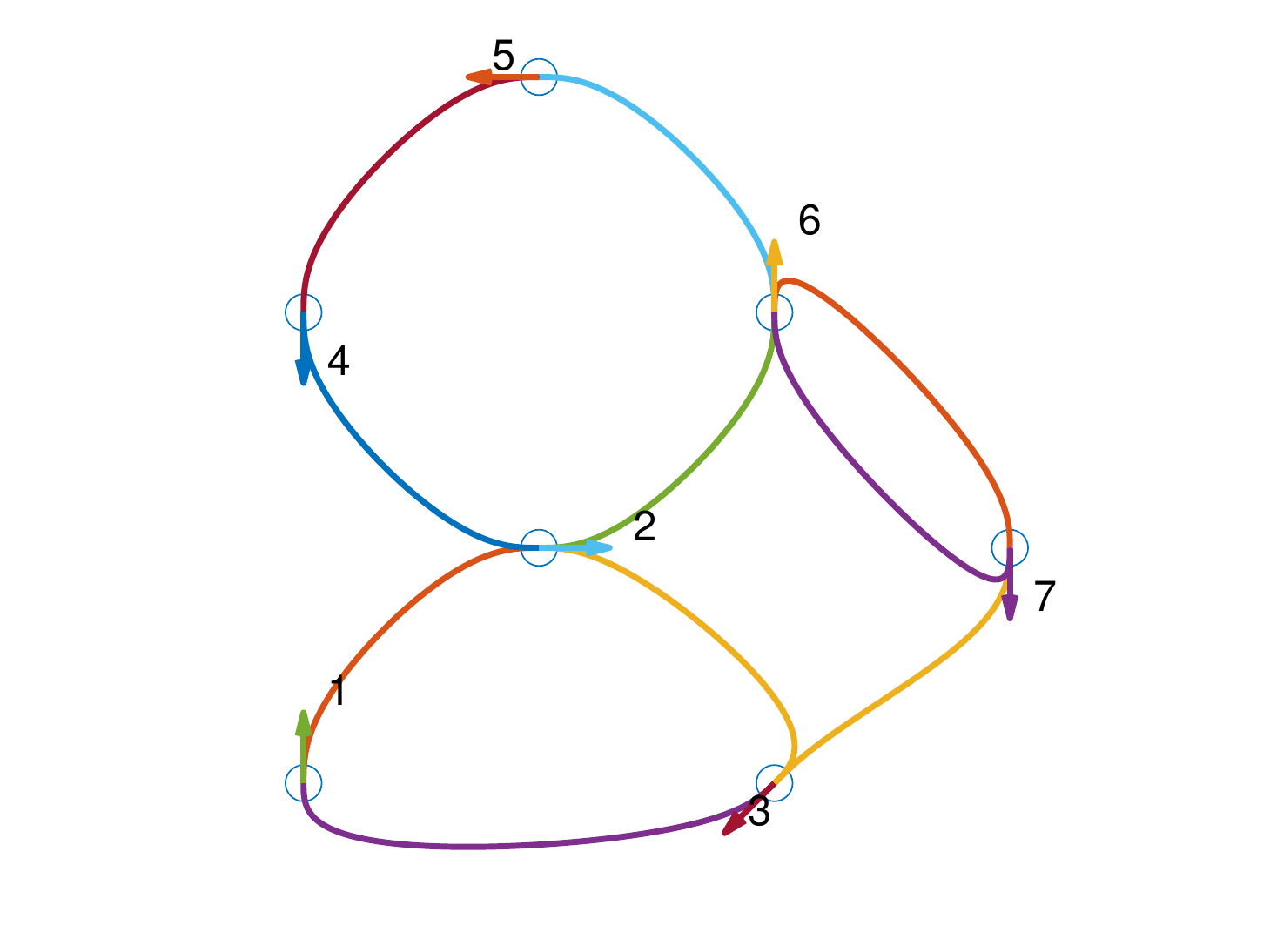}
\caption{Layout with 7 positions.}
\label{fig:layoutC}
\end{figure}

\begin{figure}
\centering
\digraph[scale=0.5]{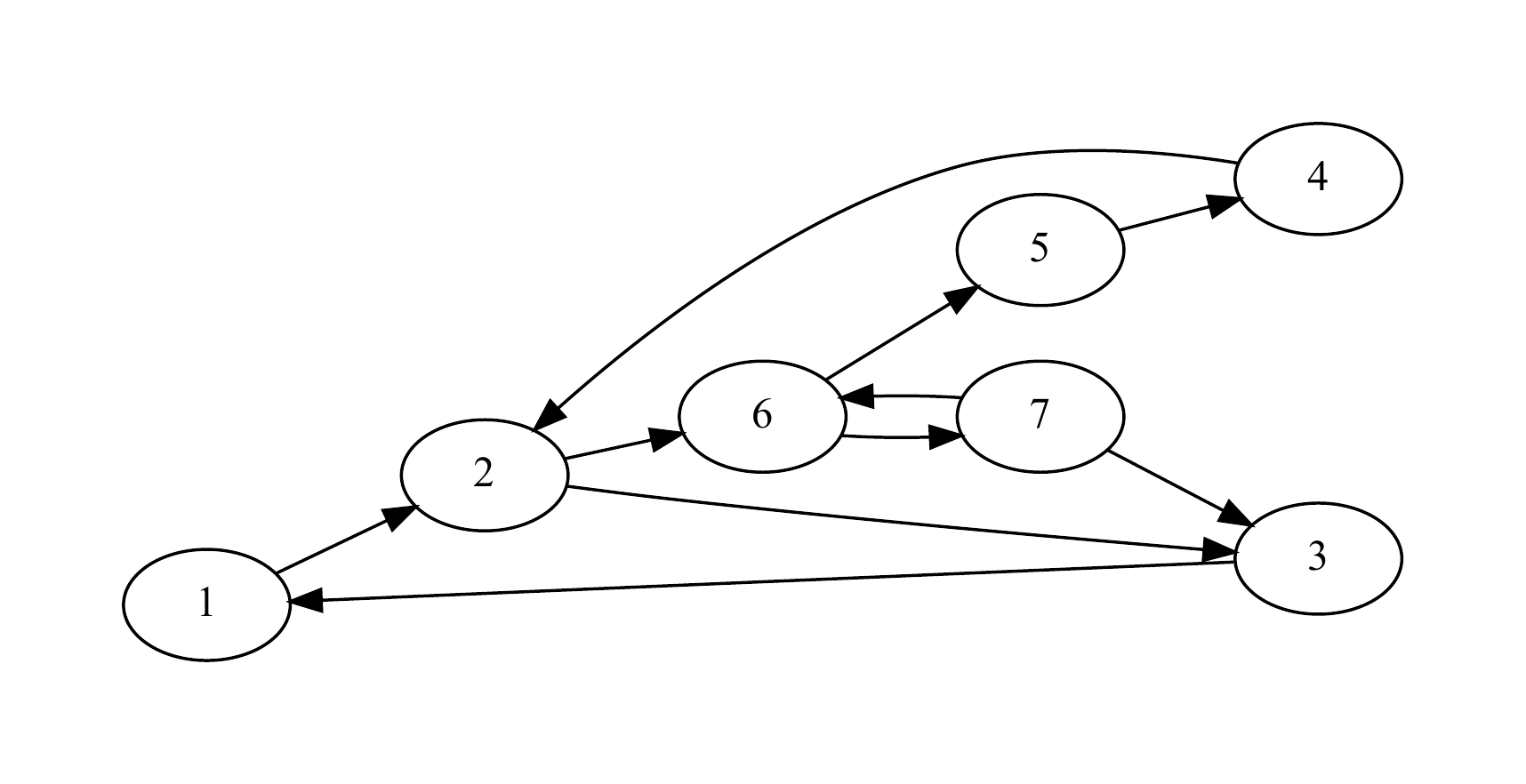}{
=LR;
1 -> 2 -> 3 -> 1;
2->6->5->4->2;
6->7->3;
7->6;
}
\caption{Directed graph associated to the setting in Figure~\ref{fig:layoutC}.}
\label{fig:sectionC}
\end{figure}

Before defining BASP in a formal way, we present an example.
Consider the setting represented in Figure~\ref{fig:layoutC}.
Here, the circles represent the positions of $7$ AGV
configurations, while the arrows represent the associated orientation angles. For instance, each configuration can be an operating point
useful to the management of an automated warehouse.  It may be a position along the racks, to insert or retrieve
packages from the shelves, a position at the end of the production lines, 
to pickup finished products, or some intermediate location, used for routing.
These configurations are connected by 10 fixed directed paths.
We can associate a directed graph to this setting, reported in
Figure~\ref{fig:sectionC}. Namely, each configuration corresponds to
a vertex and each path to a directed arc. We associate to each path
bounds on maximum and minimum velocity and acceleration, that may
depend on the arc-length position along the path, following the
procedure presented in Section~\ref{sec_assigned_path}. Roughly
speaking, solving BASP consists in finding the time-optimal motion
from a source to a destination configuration. This requires finding
both the geometrical path (i.e., the optimal sequence of directed arcs) and the
time-optimal speed law along this path that satisfied the constraints
associated
to each travelled arc. Note that, once the path is known, this last task can be
done with the method presented in Section~\ref{def:defQ}.
 
We now present BASP problem in more general terms. Let us consider a directed graph $\Gbb = (\Vbb, \Ebb)$, with $\Vbb =
\{\sigma_1, \ldots, \sigma_N\}$.
Each node $\sigma_i$, $i \in \{1,\ldots, N\}$, represents an operating point
$Q_i \in \mathbb{R}^2$.

Each arc $\theta=(\sigma_i,\sigma_j) \in \Ebb$ represents a fixed directed
path between two operating points and is associated to an arc-length
parameterized path $\gamma_{\theta}$ of length $\ell(\theta)$, such that
$\gamma_{\theta}(0)=Q_i$ and $\gamma_{\theta}(\ell(\theta))=Q_j$.


In the following, we denote the set of all possible paths on $\Gbb$
simply by $P$.
Similarly, for $s,f \in \Vbb$, we denote by $P_s$ the subset of $P$
consisting in all paths starting from $s$ and by $P_{s,f}$ the subset
of $P$ consisting in all paths starting from $s$ and ending in $f$.
We extend this definition to subsets of $\Vbb$, that is, if $S,F
\subset \Vbb$, we denote by $P_{S,F}$ the set of all paths starting
from nodes in $S$ and ending in nodes in $F$.

Given a path $p = \sigma_{1} \cdots \sigma_{m}$, its length $\ell(p)$
is defined as the sum of the lengths of its individual arcs, that is,
\[
\ell(p) = \sum_{i=1}^{m-1} \ell(\sigma_i,\sigma_{i+1}).
\]


To setup our problem, we need to associate four real-valued functions
to each edge $\theta \in \Ebb$: the maximum and minimum allowed
acceleration and squared speed.
The domain of each function is the arc-length coordinate on path
$\gamma_\theta$.
Then, given a specific path $p$, we are able to define a speed optimization
problem of form~\eqref{eqn_problem_pr_wg} by considering the constraints
associated to the edges that compose $p$.
We define the set of edge functions as
\[
\mathcal{E}=\{\varphi: \Ebb \times \Real^+ \to \Real\}.
\]
If $\varphi \in \mathcal{E}$, $\theta \in \Ebb$, $\lambda \in \Real^+$,
$\varphi(\theta,\lambda)$ denotes the value of $\varphi$ on edge
$\theta$ at position $\lambda$.
Note that $\varphi(\theta,\lambda)$ will be relevant only for $\lambda \in
[0,\ell(\theta)]$.
Given a path $p = \sigma_1 \cdots \sigma_m$, we associate to
$\varphi \in \mathcal{E}$ a function $\varphi_p:[0,\ell(p)] \to \Real$ in the
following way.
Define functions $\Theta: [0,\ell(p)] \to \mathbb{N}$, $\Lambda:[0,\ell(p)] \to \Real$
such that $\Theta(\lambda)=\max\{i \in \mathbb{N} \mid
\ell(\sigma_1\cdots \sigma_i)\leq \lambda\}$ and $\Lambda(\lambda) =
\ell(\sigma_1 \cdots \sigma_{\Theta(\lambda)})$.
In this way, $\Theta(\lambda)$ is such that $\theta(\lambda) =
(\sigma_{\Theta(\lambda)},\sigma_{\Theta(\lambda)+1})$ is the edge that
contains the position at arc-length $\lambda$ along $p$ and $\Lambda(\lambda)$
is the sum of the lengths of all arcs up to node $\sigma_{\Theta(\lambda)}$ in $p$.
Then, we define $\varphi_{p}(\lambda)=\varphi(\theta(\lambda),
\lambda-\Lambda(\lambda))$.

Given $\hat \mu^+,\hat \mu^-,\hat \alpha^+,\hat \alpha^- \in \mathcal{E}$ and path $p \in P$,
let $\BBb = (\hat \mu^-,\hat \mu^+,\hat \alpha^-,\hat
\alpha^+)$. Assume $(\forall \theta \in \Ebb) \hat \mu^+(\theta,\cdot) \in
Q_{\hat \alpha^-(\theta,\cdot),\hat \alpha^+(\theta,\cdot)}$ and define
\[
T_\BBb(p) = \min_{w \in W^{1,\infty}\left([0,s_f]\right)} \Psi(w),
\]
as the solution of Problem~\eqref{eqn_problem_pr_wg} along path $p$
with $\mu^-=\hat \mu^-_p$, $\mu^+=\hat \mu^+_p$, $\alpha^-=\hat
\alpha^-_p$, $\alpha^+=\hat \alpha^+_p$.
In this way, $T_\BBb(p)$ is the minimum-time required to traverse path
$p$, respecting the speed and acceleration constraints defined in $\BBb$.
We set $T_\BBb(p)=+\infty$ if Problem~\eqref{eqn_problem_pr_wg} is not
feasible. 

The following is the main problem discussed in this paper.
\begin{prblm}[Bounded Acceleration Shortest Path Problem (BASP)]
  \label{prob_BASP}
Given a graph $\Gbb = (\Vbb,\Ebb)$, $\mu^+,\mu^-,\alpha^-,\alpha^+ \in
\mathcal{E}$, $\BBb=(\mu^+,\mu^-,\alpha^-,\alpha^+)$, $s \in \Vbb$, and
$F \subset \Vbb$, find
\[
\min_{p \in P_{s,F}} T_{\BBb}(p).
\]
\end{prblm}

In other words, we want to find the path $p$ that minimizes the transfer
time between source node $s$ and a destination node in $F$, taking into account
bounds on speed and accelerations on each traversed arc (represented by arc
functions $\mu^+,\mu^-,\alpha^-,\alpha^+$).
The following properties are a direct consequence of the definition of
$T_\BBb(p)$.
\begin{prpstn}\label{prop_prop_T}
The following properties hold:
\hfill
\begin{itemize}
\item[i)] If $p_1,p_2 \in P$ are such that $p_1 p_2 \in P$, then
\[
T_\BBb(p_1 p_2) \geq T_\BBb(p_1) + T_\BBb(p_2).
\]

\item[ii)] If $\BBb=(\mu^+,\mu^-,\alpha^-,\alpha^+), \hat \BBb=(\hat \mu^+,\hat
\mu^-,\hat \alpha^-,\hat \alpha^+)$ are such that
$(\forall \theta \in \Ebb)$ 
$ (\forall \lambda \in [0,\ell(\theta)])$
$[\mu^-(\theta,\lambda),\mu^+(\theta,\lambda)]
\subset [\hat \mu^-(\theta,\lambda),\hat \mu^+(\theta,\lambda)]$ and
$[\alpha^-(\theta,\lambda),$\\ $\alpha^+(\theta,\lambda)]
\subset [\hat \alpha^-(\theta,\lambda),\hat \alpha^+(\theta,\lambda)]$,
then $(\forall p \in P)$
\[
T_\BBb(p) \geq T_{\hat \BBb}(p).
\]
\end{itemize}
\end{prpstn}
 In particular, the first property states that the minimum
time for travelling the composite path $p_1 p_2$ is greater or equal
to sum of the times needed for travelling $p_1$ and $p_2$ separately. In fact, in
the first case, the speed must be continuous when passing from $p_1$
to $p_2$ (due to the acceleration bounds), but this constraint does
not need to be satisfied when the speed profiles for $p_1$ and $p_2$
are computed separately.

\section{Complexity}
We discuss the complexity of a simplified version of
Problem~\ref{prob_BASP}, in which maximum and
minimum acceleration and speed are constant on each arc.

\begin{prblm}[Bounded Acceleration Shortest Path Problem with
constant bounds (BASP-C)]
\label{prob_BASP-c}
Solve Problem~\ref{prob_BASP} with the additional assumption that
there exist functions $\alpha^-,\alpha^+,\mu^-,\mu^+: \Ebb \to \Real$
such that,
$(\forall \theta \in \Ebb)\ (\forall \lambda \in \Real^+)\ \alpha^-(\theta,
\lambda) = \alpha^-(\theta), \alpha^+(\theta,\lambda)=\alpha^+(\theta),
\mu^-(\theta,\lambda)=\mu^-(\theta), \mu^+(\theta,\lambda)=\mu^+(\theta)$.
\end{prblm}
We will show that BASP-C is NP-hard, which implies
that the more general BASP is also NP-hard.

A special case of BASP-C is the classical Shortest Path (SP) problem,
where a distance/time $d(\theta)$ is associated to each edge and a minimum
distance/time path from source node $s$ to destination node $f$ is
searched for.
This is the special case when $\alpha^+(\theta)=+\infty$ and $\alpha^-(\theta)=
-\infty$ for all edges $\theta \in \Ebb$.
In this case, speed can be changed instantaneously, so that we can
run along each edge at the maximum allowed speed along that edge,
so that $d(\theta)=\frac{\ell(\theta)}{\mu^+(\theta)}$.
The classical SP problem is known to be solvable in polynomial time,
e.g., by Dijkstra's algorithm. 
BASP-C can be viewed as a generalization of the
SP problem, but, differently from SP, we prove that BASP-C is NP-hard.
The following proposition
characterizes the complexity of Problem~\ref{prob_BASP-c}.
\begin{prpstn}
  \label{prop_NP_hard}
Problem BASP-C is NP-hard.
\end{prpstn}
\begin{proof}
See Appendix \ref{sec:proofNPhard}.
\end{proof}
As said, this implies that also BASP is NP-hard.
However, we also prove that, under additional assumptions, BASP admits a pseudo-polynomial algorithm, i.e., an algorithm running in polynomial time with respect to the values of the input but not with respect to the number of bits required to represent them.
\begin{prpstn}
\label{prop_pseudo_poly}
Let us assume that the maximum and minimum acceleration along each
arc are fixed values. W.l.o.g., we assume that $\alpha^+(\theta)=+1$ and $\alpha^-(\theta)=-1$
for each $\theta \in \Ebb$.
Moreover, we also assume that all lengths $\ell(\theta)$ are positive integer values.
Then, BASP admits a pseudo-polynomial time algorithm.
\end{prpstn} 
\begin{proof}
See Appendix \ref{sec:proofpseudopoly}
\end{proof}

\section{The  $k$-BASP problem}
\label{sec_kBASP}

As stated in Proposition~\ref{prop_NP_hard}, BASP is NP-hard.
In the previous section we commented that SP can be viewed as a special case of BASP. In fact, also BASP can be viewed as an SP problem but defined on a different graph with respect to the original one. More precisely, here we introduce some restrictions on parameters $\BBb$ that
allow reducing BASP to a standard SP problem on an
extended graph, that can be solved by Dijkstra's algorithm. Let $p \in P$, define
\[
\ell^+(p)=\min\left\{\left\{\lambda \in [0,\ell(p)] \mid \int_0^\lambda
\alpha_p^+(q) dq = \mu_p^+(\lambda)\right\}, +\infty\right\},
\]
\[
\ell^-(p)=\max\left\{\left\{\lambda \in [0,\ell(p)] \mid
-\int_\lambda^{\ell(p)} \alpha_p^-(q) dq =
\mu_p^+(\lambda)\right\},-\infty\right\}.
\]

In this way, $\ell^+(p)$ is the smallest value of $\lambda \in [0,\ell(p)]$
for which the solution of $F$ in~\eqref{F_def}, with $\alpha^+=\alpha^+_p$, starting from initial
condition $F(0)=0$, reaches the squared speed upper bound $\mu^+(\lambda)$.
Note that $\ell^+(p)=\infty$ if no such value of $\lambda$ exists.
Similarly, $\ell^-(p)$ is the largest value of $\lambda \in [0,\ell(p)]$ for
which the solution of $B$ in~\eqref{B_def}, with
$\alpha^-=\alpha^-_p$, starting from initial condition
$B(\ell(p))=0$, reaches $\mu^+(\lambda)$.
Again, $\ell^-(p)=-\infty$ if no such value of $\lambda$ exists.
Note that if $p,t,pt \in P$, $\ell^+(pt) \leq \ell^+(p)$ and $\ell^-(pt)
\geq \ell^-(p)$ (actually, equalities hold if the values are all
finite). Finally, we define
\begin{equation}
\label{eqn_def_of_K}
  K(\BBb)=\min\{k \in \mathbb{N} \mid (\forall p \in P_s)\ |p|\geq k \Rightarrow
\ell^+(p) \leq \ell^-(p)\}.
\end{equation}

\begin{figure}
  \begin{center}
  \begin{tikzpicture}
    \tikzstyle{every node}=[circle, draw, fill=black,
                        inner sep=0pt, minimum width=4pt]
  \node [label=$s$] (s)                    {};
  \node         (1) [right=2cm of s,label=$1$] {};

  \node         (2) [right=2cm of 1,label=$2$] {};
  \node     (f) [right= 2cm of 2, label=$f$] {};

  \path [->] (s) edge (1); 
  \path[->] (1) edge (2);
  \path[->] (2) edge (f);
 \end{tikzpicture} 
    \caption{Simple graph with source node $s$ and final node $f$.}
    \label{fig:example_k}
\end{center}
\end{figure}
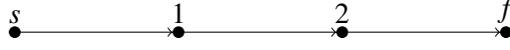

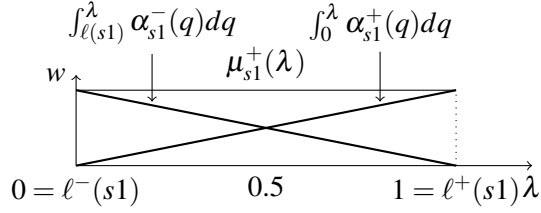
\begin{figure}
  \begin{center}
    \begin{tikzpicture}[y=1cm, x=5cm]

      \draw[->] (0,0) -- (1.2,0) node[anchor=north] {$\lambda$};
\draw	(0,0) node[anchor=north] {$0=\ell^-(s1)$}
		(0.5,0) node[anchor=north] {0.5}
		(1,0) node[anchor=north] {$1=\ell^+(s1)$};

\draw[->] (0,0) -- (0,1.2) node[anchor=east] {$w$};
                
                \draw [thick] (0,0)--(1,1);
\draw (0.8,1.9) node (int+) {$\int_0^\lambda \alpha_{s1}^+(q) dq$}; 
                \draw [thick] (0,1)--(1,0);
\draw (0.2,1.9) node (int-){$\int_{\ell(s1)}^\lambda \alpha_{s1}^-(q) d q$}; 
\draw (0,1) --  node [above] {$\mu_{s1}^+(\lambda)$} (1,1);
\draw [dotted] (1,0) -- (1,1);
\draw[->] (int+)--(0.8,0.85);
\draw[->] (int-)--(0.2,0.85);
 
\end{tikzpicture}
 \caption{Computation of $\ell^+(s1)=1$, $\ell^-(s1)=0$.}
    \label{fig:example_k_1}
\end{center}
\end{figure}

\begin{figure}
  \begin{center}
    \begin{tikzpicture}[y=1cm, x=3cm]

      \draw[->] (0,0) -- (2.2,0) node[anchor=north] {$\lambda$};
\draw	(0,0) node[anchor=north] {0}
		(1,0) node[anchor=north east] {$\ell^+(s12)=1$}
		(4/3,0) node[anchor=north west] {$\frac{4}{3}=\ell^-(s12)$};
		(2,0) node[anchor=north] {2};

\draw[->] (0,0) -- (0,1.2) node[anchor=east] {$w$};
                
\draw [thick] (0,0)--(1,1);
\draw (0.8,1.4) node (int+) {$\int_0^\lambda \alpha_{s12}^+(q) dq$}; 
\draw [thick] (2,0)--(4/3,2/3);
\draw (1.8,1.4) node (int-){$\int_{\ell(s12)}^\lambda \alpha_{s12}^-(q) d q$}; 
\draw (0,1) --  (1,1);
\draw (1,2/3) node [below] {$\mu_{s12}^+(\lambda)$}  -- (2,2/3);
\draw(1,1)--(1,2/3);
\draw [dotted] (1,0) -- (1,2/3);
\draw [dotted] (2,0) -- (2,2/3);
\draw [dotted] (4/3,0) -- (4/3,2/3);

\draw[->] (int+)--(0.8,0.85);
\draw[->] (int-)--(1.8,0.25);
 
\end{tikzpicture} 
    \caption{Computation of $\ell^+(s12)=1$, $\ell^-(s12)=\frac{4}{3}$.}
    \label{fig:example_k_2}
\end{center}
\end{figure}
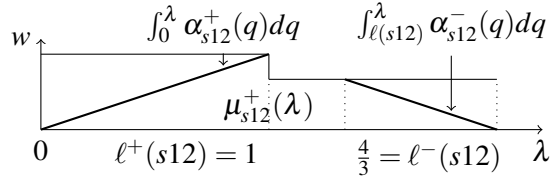

We call $k$-BASP any instance of Problem~\ref{prob_BASP} that
satisfies $K(\BBb) \leq k$.
For instance, consider the simple graph depicted in
Figure~\ref{fig:example_k}. Here, $\Vbb=\{s,1,2,f\}$,
$\Ebb=\{(s,1),(1,2),(2,f)\}$, $(\forall \theta \in \Ebb)\
\alpha^-(\theta)=-1$, $\alpha^+(\theta)=1$, $\mu^-(\theta)=0$, $\ell(\theta)=1$, moreover
$\mu^+((s,1))=1$, $\mu^+((1,2))=\frac{2}{3}$, $\mu^+((2,f))=1$.
In this case, $P_s=\{s,s1,s12,s12f\}$.
Moreover, $K(\BBb)>2$, since $\ell^+(s1) =1 > 0=\ell^-(s1)$ as
reported in Figure~\ref{fig:example_k_1}.
Further, $\ell^+(s12) < \ell^-(s12)$ and
$\ell^+(12f) < \ell^-(12f)$ and $s12,12f$ are the only paths of length
$3$.
Figure~\ref{fig:example_k_2} shows the computation of $\ell^+(s12)$
and $\ell^-(s12)$, the computation of $\ell^+(12f)$
and $\ell^-(12f)$ is analogous. Hence, in this example,
$K(\BBb)=3$. 

Note that $K(\BBb)-1$ represents the  maximum number of vertices of a path that can be
traveled with a speed profile of maximum acceleration, followed by
one of maximum deceleration, starting and ending with zero speed, without violating the maximum speed constraint. 
The following expression provides a simple upper bound on $K(\BBb)$
\begin{equation}
  \label{eqn_bound_k}
K(\BBb) \leq 1+ \left\lceil 2 \max_{\theta \in \Ebb}
\frac{\displaystyle \max_{\lambda \in [0,\ell(\theta)]} \mu^+(\theta,\lambda)}
{\displaystyle \min_{\lambda \in [0,\ell(\eta)]}
 \left(\alpha^+(\theta,\lambda) \wedge |\alpha^-(\theta,\lambda)|\right) \ell(\theta)}\right\rceil.
\end{equation}

Note that $K(\BBb)=1$ only if $\alpha_-=-\infty$ and
$\alpha^+=+\infty$, that is, if we do not consider acceleration bounds.
We will present an algorithm that solves
$k$-BASP in polynomial time complexity with respect to $|\Vbb|$
and $|\Ebb|$.
However, note that the complexity is exponential with respect to $k$,
so that a correct estimation of $K(\BBb)$ si critical.  In general, bound~\eqref{eqn_bound_k}
overestimates $K(\BBb)$. In section~\ref{sec_adap_k} we will present a simple
method for correctly estimating $K(\BBb)$.


Define $\Suff_k: P \to \Vbb_k$ such that, if $|p| \leq k$, $\Suff_k (p)=p$
and if $|p| > k$, $\Suff_k (p)$ is the suffix of $p$ of length $k$. 
Function $\Suff_k$ allows to introduce a partially defined transition
function $\Gamma: \Vbb_k \times \Vbb \to \Vbb_k$ by setting
\[
\Gamma(r,\sigma) =
\begin{cases}
\Suff_k(r \sigma), & \textrm{if } r \sigma \in P \\
\textrm{is not defined}, & \textrm{if } r \sigma \notin P.
\end{cases}
\]

We recall that $\Vbb_k$ represents the subset of language $\Vbb^*$
composed of strings with maximum length $k$, including the empty string $\epsilon$.



Define the incremental cost function $\eta: P_s \times \Vbb \to
\Real^+$ such that, for $p \in P_s$ and $\sigma \in \Vbb$,
\[
\eta (p,\sigma) =
\begin{cases}
T_{\BBb}(p \sigma)-T_{\BBb}(p), & \textrm{if } p \sigma \in P_s \\
+\infty, & \textrm{otherwise}.
\end{cases}
\]
                             
In other words, $\eta(p,\sigma)$ is the difference between the
minimum-time required for traversing $p\sigma$ and the minimum-time
required for traversing $p$.
For simplicity of notation, from now on we will denote $T_{\BBb}$
simply as $T$. The following proposition shows that the incremental cost is always
strictly positive.
\begin{prpstn}
  \label{prop_positive}
$\eta(p,\sigma) \geq T(\sigma)$.
\end{prpstn}
\begin{proof}
By i) of Proposition~\ref{prop_prop_T},
$T(p \sigma)\geq T(p)+T(\sigma)$.
\end{proof}

The following property, whose proof is presented in the Appendix,
plays a key role in the solution algorithm. 
\begin{prpstn}
\label{prop_lpm}
Let $p_1,p_2,t \in P$ be such that $p_1t, p_2t \in P$ and
$\ell^+(t) \leq \ell^-(t)$, then $(\forall \sigma \in \Vbb)$
\[
T(p_1 t \sigma)-T(p_1 t)= T(p_2 t \sigma)-T(p_2 t).
\]
\end{prpstn}

The following is a direct consequence of Proposition~\ref{prop_lpm}.
It states that, given $p
\in P$ and $\sigma \in \Vbb$, the
incremental cost $\eta(p,\sigma)$ does not depend on the complete
path $p$, but only on $\Suff_k(p)$ (its last $k$ symbols).

\begin{prpstn}
  \label{prop_suffix}
If $K(\BBb) \leq k$ and $p, p' \in P$ are such that
$\Suff_k(p)=\Suff_k(p')$, then $(\forall \sigma \in \Vbb)$
\[
\eta(p,\sigma)=\eta(p',\sigma)\,.
\]
\end{prpstn}

Define function $\hat \eta: \Vbb_k \times \Vbb \to \Real^+$, such that
$\hat \eta(r,\sigma)=\eta(p,\sigma)$ where $p \in P$ is any path
such that $r=\Suff_k(p)$.
We set $\hat \eta(r,\sigma)=+\infty$ if such path does not exist.
Note that function $\hat \eta$ is well-defined by Proposition~\ref{prop_suffix},
being $\eta(p,\sigma)$ identical among all paths $p$ such that $r=\Suff_k(p)$.
In particular, Proposition~\ref{prop_suffix} holds for $p'=\Suff_k(p)=r$,
so that we can compute $\hat \eta$ as
\[
\hat \eta(r,\sigma)=\eta(r,\sigma).
\]
In the following, since $\hat \eta$ is the restriction of $\eta$ on
$\Vbb_k \times \Vbb$, we will denote $\hat \eta$ simply by $\eta$.

The value $k$ can be viewed as the amount of memory required to solve
the problem: once a node is reached, the optimal path from such node
to the target one depends on the last $k$ visited nodes. If $k=1$, it only depends on the current node itself (i.e., no memory is required). This is the situation with the classical SP problem. More generally, $k>1$, so that the optimal way to complete  the path does not only depend on the current node, but also on the sequence of $k-1$ nodes visited before reaching it.

Define function $V:\Vbb_k \to \Real$ as
\begin{equation}
  \label{eqn_for_v}
V(r)=\min_{p \in P_s \mid \Suff_k p=r} T_{\BBb}(p).
\end{equation}

Note that the solution of BASP corresponds to
$\min_{r \in \Vbb_k \mid \vec r \in F} V(r)$ (we recall that $\vec r$ is
the last vertex of $r$).
For $r \in \Vbb_k$, define the set of predecessors of $r$ as
$\Prec(r)=\{\bar{r} \in \Vbb_k\mid r=\Gamma(\bar{r},\vec{r})\}$. The following proposition presents an expression for $V(r)$ that holds if condition $\ell^{+}(r') \leq \ell^{-}(r')$ is satisfied for all predecessors $r'$ of $r$.
\begin{prpstn}
\label{prop_for_V}
Let $r \in \Vbb_k$, if $(\forall r' \in \Prec(r))\ \ell^+(r') \leq \ell^-(r')$, then 
\begin{equation}
  \label{eqn_for_V}
V(r) = \min_{r' \in \Prec(r)}
\{V(r')+ \eta(r',\vec{r})\}.
\end{equation}
\end{prpstn}

\begin{proof}
 \[
\begin{gathered}
V(r)=\min_{p \in P_s \mid \Suff_k p=r} T(p)=\\
\min_{q\in P_s \mid \Suff_k q \vec{r}=r} \{T(q \vec{r}) - T(q) + T(q)\}= \\
\min_{q\in P_s \mid \Suff_k q \vec{r}=r} \{T(q)+ T((\Suff_k q) \vec{r}) -
T(\Suff_k q) \}=\\
\min_{q\in P_s \mid \Suff_k q \vec{r}=r} \{T(q)+ \eta(\Suff_k q, \vec{r})\}
=\\ \min_{q \in P_s, r' \in \Prec(r) \mid \Suff_k q =r'}
\{T(q)+ \eta(r',\vec{r})\}
=\\ \min_{r' \in \Prec(r)}
\{V(r')+ \eta(r',\vec{r})\},
\end{gathered}
\]
where we used the facts that $T(q \sigma) - T(q) =T(\Suff_k q
\sigma)- T(\Suff_k q)$, by Proposition~\ref{prop_lpm},
and that $q \in P_s$ is such that $\Suff_k q \vec{r}=r$ if and only if
$\Suff_k q \in \Prec (r)$.
\end{proof}

\begin{figure}
\centering
\digraph[scale=0.5]{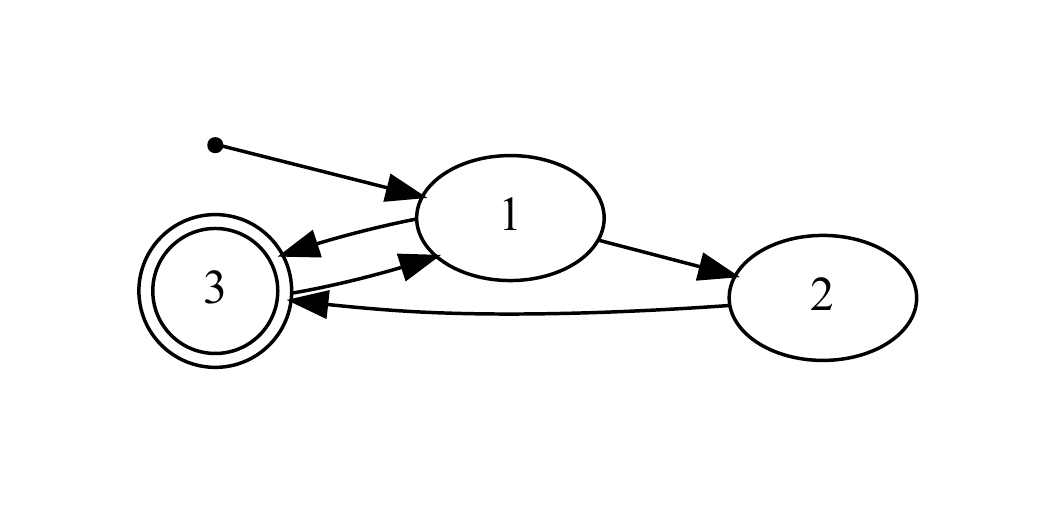}{
  rankdir=LR;
    3 [shape = doublecircle]; 
    init [label="", shape=point];
   init-> 1 -> 2 -> 3 -> 1;
  1 -> 3;
}
\digraph[scale=0.5]{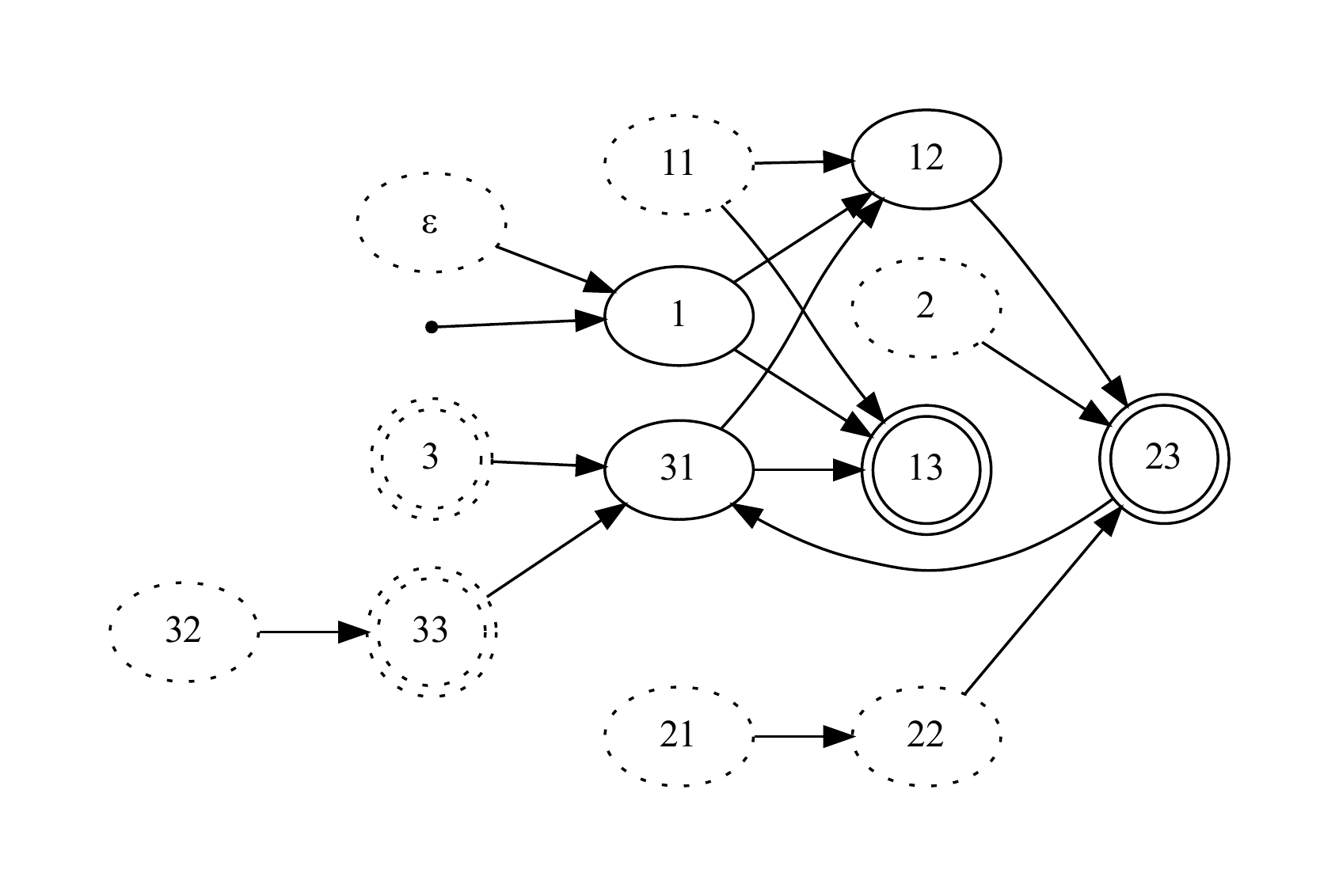}{
  rankdir=LR;
  3 [shape = doublecircle];
  13 [shape = doublecircle];
  23 [shape = doublecircle];
  33 [shape = doublecircle];
  epsi [label=<&epsilon;>,style=dotted];
  2 [style = dotted];
  3 [style = dotted];
  init [label="", shape=point];
  init->1;
  epsi->1;
  1->12;
  1->13;
  2->23;
  3->31;
  12->23;
  23->31;
  31->12;
  31->13;
  11 [style = dotted];
  22 [style = dotted];
  33 [style = dotted];
  32 [style = dotted];
  21 [style = dotted];

  11->13;
  11->12;
  22->23;
  33->31;
  32->33;
  21->22;  
}
\caption{A graph and the corresponding extended graph for $k=2$.}
\label{fig:example_ext_graph}
\end{figure}

As a consequence of Proposition~\ref{prop_for_V}, if $(\forall r \in
\Vbb_k)\ \ell^+(r) \leq \ell^-(r)$, $V(r)$ corresponds
to the length of the shortest path from $s$ to $r$ on the extended
directed graph $\tilde \Gbb=(\tilde \Vbb,\tilde \Ebb)$, where $\tilde \Vbb=
\Vbb_k$ and $(r_1,r_2) \in \tilde \Ebb$ if $r_2=\Gamma(r_1,\vec{r_2})$ is
defined,  in this case its length is $\eta(r_1,\vec{r_2})$. 
The upper part of Figure~\ref{fig:example_ext_graph} shows
a graph consisting of $3$ nodes. Node $s=1$ is the source (indicated
by the entering arrow) and the double border shows the final node $F=\{3\}$.
The lower part of Figure~\ref{fig:example_ext_graph} represents
the corresponding extended graph, obtained for $k=2$, consisting of
$13$ nodes (the cardinality of $\Vbb_2$).
Note that some of the nodes are unreachable from the initial state,
these are represented with dotted edges.

Solving $k$-BASP corresponds to finding a minimum-length path
on $\tilde \Gbb$ that connects node $s \in \Vbb_k$ to $\hat F=\{r \in
\Vbb_k \mid \vec{f} \in F\}$. Note that the set of final states for the
extended graph $\hat F$
contains all paths $p \in \Vbb_k$ that end in an element of $F$.
In the extended graph reported in
Figure~\ref{fig:example_ext_graph}, this corresponds to finding a
minimum-length path from starting node $1$ to one of the final nodes
$3$, $13$, $23$, $33$. Note that the unreachable nodes play no role in
this procedure.
We can find a minimum-length path by Dijkstra's algorithm applied on $\tilde \Gbb$, leading
to the following complexity result.

%


\begin{prpstn}
  \label{prop_complexity}
$k$-BASP can be solved with complexity
$O(|\Vbb|^{k-1}|\Ebb| + (|\Vbb|^{k} \log |\Vbb|^{k}))$. 
\end{prpstn}

\begin{proof}
Dijkstra's algorithm has time complexity $O(|E|+|V| \log |V|)$, where
$|E|$ and $|V|$ are the cardinalities of the edge and vertex sets.
In our case,
$|V|=|\tilde \Vbb|=|\Vbb_k|=\sum_{i=0}^k |\Vbb|^i=O(|\Vbb|^k)$,
$|E|=|\tilde \Ebb| \leq |\Vbb_{k-1} \Ebb|=O(|\Vbb|^{k-1} |\Ebb|)$,
which imply the thesis.
\end{proof}

The following remark establishes again that SP can be viewed as a special case of BASP when no bound on the acceleration is imposed.
\begin{rmrk}
  \label{remark_1_basp}
If $(\forall \sigma \in \Vbb)\ (\forall \lambda)\ \alpha^-(\sigma,\lambda)=-\infty$,
$\alpha^+(\sigma,\lambda)=+\infty$, then $K(\BBb)=1$.
The resulting $1$-BASP reduces to a standard SP problem on graph $\Gbb$
and can be solved with time complexity $O(|\Ebb|+|\Vbb| \log |\Vbb|)$.
\end{rmrk}

\section{Adaptive A$^*$ algorithm for $k$-BASP}
The computation method based on Dijkstra's algorithm on the extended
graph $\tilde \Gbb$, presented in the previous section, has two main
disadvantages.
First, the extended graph has $\sum_{j = 1}^k |\Vbb|^j$
nodes, so that the time required
by Dijkstra's algorithm grows exponentially with $k$.
We will show that it is possible to mitigate this problem and reduce the
number of visited nodes by using A$^*$ algorithm with a suitable heuristic.
Second, the estimation of $k=K(\BBb)$ from its definition is not an easy task.
We will show that it is quite easy to adaptively find the correct value of $k$
by starting from $k=2$ and increasing $k$ if needed.

\subsection{Upper bounds on $T_{\BBb}(p)$}
To implement the A$^*$ algorithm, we need to define a heuristic function
$h:\Vbb_k \to \Real$, such that, for $r \in \Vbb_k$, $h(r)$ is a lower
bound on $\min_{p \in P_{\vec{r},\hat F}} T(p)$, that is, the minimum time
needed for traveling from $\vec{r}$ to a final state in $\hat F$.
In general, we can compute lower bounds for BASP  by relaxing the
acceleration constraints $\alpha^-$, $\alpha^+$.
Namely, let $\hat \BBb$ be a parameter set obtained by relaxing acceleration
constraints in $\BBb$.
Then, if $K(\hat \BBb) < K(\BBb)$, by Proposition~\ref{prop_complexity},
the solution of BASP for parameters $\hat \BBb$ can be computed with a lower
computational time than the solution with parameters $\BBb$.
In particular, we obtain a very simple lower bound by removing
acceleration bounds altogether, that is, by setting $\alpha^-=-\infty$
and $\alpha^+=+\infty$.
In this way, the vehicle is allowed to travel at maximum speed
everywhere along the path and the incremental cost function
$\eta(p,\sigma)$ is given by the time needed to travel
$\gamma_\sigma$ at maximum speed, that is:
\[
\eta(p,\sigma) = \int_{0}^{\ell(\vec{p}\sigma)}
\frac{1}{\sqrt{\mu^+((\vec{p},\sigma),\lambda)}} d\lambda.
\]


Define the heuristic $h: \Vbb_k \to \Real^+$ as
\begin{equation}
  \label{eqn_heu}
h(r)=\min_{p \in P_{\vec{r},\hat F}} T_{\hat \BBb}(p).
\end{equation}

Note that, if $\alpha^-=-\infty$
and $\alpha^+=+\infty$, $h$ corresponds to the solution of $1$-BASP and
all values of $h$ can be efficiently precomputed by Dijkstra's
algorithm (see Remark~\ref{remark_1_basp}).

The following proposition shows that $h$ is admissible and consistent,
so that A$^*$ algorithm, with heuristic $h$, provides the optimal solution of
$k$-BASP and its time-complexity is no worse than Dijkstra's algorithm
(see for instance Theorems~2.9 and~2.10 of~\cite{edelkamp2011heuristic}). 

\begin{prpstn}
Heuristic $h$ satisfies the following two properties:
\begin{itemize}
\item[i)] $(\forall r \in \Vbb_k)\ h(r) \leq \min_{q \in P_{\vec{r},f}} T_{\BBb}(q)$
(admissibility).

\item[ii)] $(\forall r \in \Vbb_k)\ (\forall \sigma \in \Vbb)\
  h(r) \leq \eta(r,\sigma) + h(\Gamma(r,\sigma))$ (consistency).
\end{itemize}
\end{prpstn}

\begin{proof}
i)  $h(r)=\min_{p \in P_{\vec{r},f}} T_{\hat \BBb}(p) \leq \min_{q
\in P_{\vec{r},f}} T_{\BBb}(q)$, since $\hat \BBb$ is a relaxation of $\BBb$.

ii) $h(r)= \min_{p \in P_{\vec{r},f}} T_{\hat \BBb}(p) \leq T_{\hat
  \BBb}(\sigma)+\min_{p \in P_{\sigma,f}} T_{\hat \BBb}(p)
\leq T_{\BBb}(\sigma)+\min_{p \in P_{\sigma,f}} T_{\hat \BBb}(p) $ $
\leq
\eta(r,\sigma) + \min_{p \in P_{\sigma,f}} T_{\hat \BBb}(p)=\eta(r,\sigma)
+ h(\Gamma(r,\sigma))$, where $ T_{\hat
  \BBb}(\sigma)\leq  T_{\BBb}(\sigma)$ by ii) of
Proposition~\ref{prop_prop_T} and $T_{\BBb}(\sigma) \leq
\eta(r,\sigma)$ by Proposition~\ref{prop_positive}.
\end{proof}

Since heuristic $h$ is admissible and consistent, A$^*$ is
equivalent to Dijkstra's algorithm, with the only difference that the
incremental cost function $\eta(r,\sigma)$ is substituted with
modified cost
\begin{equation}
  \label{eqn_tilde_eta}
  \tilde \eta(r,\sigma)=\eta(r,\sigma)+h(\Gamma(r,\sigma))-h(r)
  \end{equation}
(see Lemma~2.3 of \cite{edelkamp2011heuristic} for a complete discussion).
A description of A$^*$ algorithm can be found in literature (for
instance, see Algorithm~2.13 of~\cite{edelkamp2011heuristic}).
For the sake of completeness, we report a possible implementation.
We define a priority queue $\Qc$ that contains
open nodes, that is, nodes that have
already been generated but have not yet been visited. 
Namely, $\Qc$ is an ordered set of pairs $(r,t) \in \Vbb_k \times \Real^+$,
in which $r \in \Vbb_k$ and $t$ is a lower bound for the time
associated to the best completion of $r$ to a path arriving at a final
state.
We need to perform the following operations on $\Qc$: find its
element with the minimal $t$-value, insert a pair, and update the
queue if a node improves its $t$-value due to the discovery of a shorter path.
Accordingly, we define the following operations on $\Qc$.
Operation $\textproc{Insert}(\Qc,(r,t))$ inserts couple $(r,t)$ into
$\Qc$, operation $(r,t)=\textproc{DeleteMin}(\Qc)$ returns the first
couple of $\Qc$, that is, the couple $(r,t)$ with the minimum time
$t$, and removes this couple from $\Qc$.
Finally, operation $\textproc{DecreaseKey}(\Qc,(r,t))$ assumes that
$\Qc$ already contains a couple $(r,t')$ with $t'>t$ and substitutes
this couple with $(r,t)$.
Further, we consider three partially defined maps $\textproc{value}:
\Vbb_k \to \Real$, $\textproc{parent}:\Vbb_k \to \Vbb_k$,
$\textproc{closed}: \Vbb_k \to \{0,1\}$, such that, for $r \in \Vbb_k$,
$\textproc{value}(r)$ is the current best upper estimate of $V(r)$,
$\textproc{parent}(r)$ is the parent node of $r$ and $\textproc{closed}
(r)=1$ if node $r$ has already been visited.
Maps $\textproc{value}$, $\textproc{parent}$, and $\textproc{closed}$
can be implemented as hashtables.
For a complete discussion on A$^*$ algorithm and the data structures
involved, we refer again the reader to~\cite{edelkamp2011heuristic}.


\begin{lgrthm}[A$^*$ algorithm for k-BASP]
\label{alg_enh}
\hfill
\begin{itemize}
\item[1)] [initialization]  Set $\Qc=\{(s,h(s))\}$, $\textproc{value}(s)=0$.
  
\item[2)] [expansion]
Set $(r,t)=\textproc{DeleteMin}(\Qc)$ and set $\textproc{closed} (r)=1$.
If $\vec r \in \hat F$, then $t$ is the optimal solution and the algorithm
terminates, returning maps $\textproc{value},\textproc{parent}$.
Otherwise, for each $\sigma \in \Vbb$ for which $\Gamma(r, \sigma)$ is defined,
set $r'=\Gamma(r,\sigma)$, $t'=t+\tilde \eta (r,\sigma)$.
If $\textproc{closed}(r')=1$, go to 3). Else, if
$\textproc{value}(r')$ is undefined
$\textproc{Insert}(\Qc,(r',t'))$.
Otherwise, if $t'<\textproc{value}(r')$, set $\textproc{value}(r')=t'$,
$\textproc{parent}(r')=r$ and do $\textproc{DecreaseKey}(\Qc,(r',t'))$.

\item[3)] [loop] If $\Qc$ is not empty go back to 2),
otherwise no solution exists.
\end{itemize}
\end{lgrthm}

\begin{prpstn}
Algorithm~\ref{alg_enh} terminates and
returns the optimal solution (if it exists), with a time-complexity not higher than
Dijkstra's algorithm.
\end{prpstn}

\begin{proof}
It is a consequence of the fact that heuristic $h$ is admissible
and consistent (see, for instance, Theorems~2.9 and~2.10
of~\cite{edelkamp2011heuristic}).
\end{proof}

Note that, at the end of Algorithm~\ref{alg_enh}, $\textproc{value}(f)$
is the optimal value of $k$-BASP and the optimal path from $s$ to set $F$ can
be reconstructed from map $\textproc{parent}$.


\subsection{Adaptive search for $k$}
\label{sec_adap_k}
One possible limitation of Algorithm~\ref{alg_enh} is that estimating
$K(\BBb)$ from its definition can be difficult.
A correct estimation of $K(\BBb)$ is critical for the efficiency of the algorithm.
Indeed, if $K(\BBb)$ is overestimated, the time-complexity of the algorithm is
higher than it would be with a correct estimate.
On the other hand, if $K(\BBb)$ is underestimated, Algorithm~\ref{alg_enh}
is not correct since Proposition~\ref{prop_for_V} does not hold.
Here we propose an algorithm that adaptively find a suitable value for
$k$ in Algorithm~\ref{alg_enh}, that may be lower or equal to the true
value of $K(\BBb)$, but, in any case, allows
to find the optimal solution of BASP.
First, we define the modified cost function $W:\Vbb_k \to \Real$ as
\[
W(r)=V(r) + h(r),
\]
where $V$ is given by~\eqref{eqn_for_v} and $h$ is the heuristic given
by~\eqref{eqn_heu}.

If $(\forall r \in \Vbb_k)\ \ell^+(r) \leq \ell^-(r)$, then $W$ is the solution of
\begin{equation}
\label{eqn_for_W}
\begin{cases}
W(r) = \min_{r' \in \Prec(r)} \{W(r')+ \tilde \eta(r',\vec{r})\} \\
W(s)=h(s)
\end{cases}
\end{equation}
      
Indeed, following the same steps of the proof of Proposition~\ref{prop_for_V}
\[
\begin{gathered}W(r)=V(r)+ h(r) =\\
\min_{r' \in \Prec(r)} \{V(r') + \eta(r',\vec{r})+ h(r)+h(r')-h(r')\} \\
=\min_{r' \in \Prec(r)}
\{W(r')+ \tilde \eta(r',\vec{r})\}.
\end{gathered}
\]
Hence, $W(r)$ corresponds to the length of the shortest path from $s$ to
$r$ on $\tilde \Gbb$, with arc-length given according to $\tilde \eta$.
If condition $\ell^+(r) \leq \ell^-(r)$ is not satisfied for all $r \in
\Vbb_k$, equation~\eqref{eqn_for_W} does not hold for all $r \in
\Vbb_k$ and $W$ does not represent the solution of a shortest path
problem. However, the following proposition shows that we can still find a lower
bound $\hat W$ of $W$ that does
correspond to the solution of a shortest path problem.

\begin{prpstn}
\label{prop_V_hat}
Let $\hat W: \Vbb_k \to \Real$ be the solution of
\begin{equation}
\label{eqn_bound}
\begin{cases}
\hat W(r) = \min_{r' \in \Prec(r)} \{\hat W(r')+ \hat \eta(r',\vec{r})\} \\
\hat W(s)=0
\end{cases}
\end{equation}
where
\[
\hat \eta(r',\vec{r})=
\begin{cases}
\tilde \eta(r',\vec{r}),	& \textrm{if } \ell^+(r') \leq
\ell^-(r') \textrm{ or } |r'| < k\\
h(r)-h(r'),			& \textrm{otherwise.}
\end{cases}
\]
Then, $(\forall r \in \Vbb_k)$
\hfill
\begin{itemize}
\item[i)] $\hat W(r) \leq W(r)$.
\item[ii)] if $(\forall \bar r \in \Vbb_k \mid \hat W(\bar r) \leq \hat W(r))\,
\ell^+(\bar r) \leq \ell^-(\bar r)$, then $\hat W(r)=W(r)$.
\end{itemize}
\end{prpstn}

\begin{proof}
i) For $r \in \Vbb_k$, let $p \in P_s$ be such that $\Suff_k p \in \Prec(r)$.
If $\ell^+(\Suff_k p) \leq \ell^-(\Suff_k p)$, in view of
Proposition~\ref{prop_lpm}, $T(p
\vec{r})=T(p) + \eta (\Suff_k p,\vec{r})$, otherwise, obviously, 
$T(p\vec{r}) \geq T(p)$. Hence, in both cases, by the definition of
$\tilde \eta$ in~\eqref{eqn_tilde_eta}, $T(p \vec{r}) + h(r) \geq
T(p) + h(\Suff_k p)+ \hat \eta(\Suff_k p, \vec{r})$.
By contradiction, assume that there exists a non-empty subset $A \subset
\Vbb_k$ such that $(\forall r \in A)\ \hat W(r) >W(r)$.
Let $\bar r=\operatorname{argmin}_{\hat r \in A} W(\hat r)$, then,
\[
\begin{gathered}W(\bar r)=V(\bar r)+ h(\bar r) = \min_{p \in P_s \mid
\Suff_k p=\bar r} T(p) + h(\bar r)= \\
\min_{q \in P_s \mid \Suff_k q \in \Prec (\bar r)} T(q \vec{r}) + h(\bar r) \geq \\
\min_{q \in P_s \mid \Suff_k q \in \Prec (\bar r)} \{T(q) + h(\Suff_k(q))
+ \hat \eta(\Suff_k q,\vec{\bar r})\} = \\
\min_{r' \in \Prec (\bar r)} \{\hat W(r') + \hat \eta(r',\vec{\bar r})\} = \hat W(\bar r),   
\end{gathered}
\]
where we used the fact that $W(r')=\hat W(r')$, that follows from the
definition of $\bar r$, since the value of
$r'$ that attains the minimum is such that $W(r') < W(\bar r)$.
Then, the obtained inequality contradicts the fact that $\hat W(\bar r)> W(\bar r)$.

ii) Let $A \subset \Vbb$ be the set of values of $r
\in \Vbb$ for which ii) does not hold and, by contradiction, assume
that $A$ is not empty and let $\hat r=\textrm{argmin}_{r \in A} \hat
W(r)$. Then, by the definition of $\hat r$, it satisfies the following
two properties. First, $(\forall \bar r \in
\Vbb_k \mid \hat W(\bar r) \leq \hat W(\hat r))\,
\ell^+(\bar r) \leq \ell^-(\bar r)$, moreover $\hat W(\hat r) \neq
W(\hat r)$.

Note that, from the definitions of $\hat W$, $W(s)=\hat W(s)$. Then,
\[
\begin{gathered}
W(\hat r)=\min_{p \in P_s \mid \Suff_k p=\hat r} T(p)+ h(\hat r)= \\
\min_{q \in P_s \mid \Suff_k q \in \Prec (\hat r)}
\{T(q \vec{\hat r}) + h(\Suff_k q)-h(\Suff_k q) + h(\hat r)\} \\
= \min_{r'  \in \Prec (\hat r)} \{\hat W(r')+ \hat \eta(r',\vec{\hat
  r})\} = \hat W(\hat r),
\end{gathered}
\]
which contradicts the definition of $\hat r$.
Here, we used equation~\eqref{eqn_tilde_eta} and the fact that, since $\hat W(r')
< \hat W(\hat r)$ and by the definition of $\hat r$, $\hat W(r') =
W(r')$.
\end{proof}

Proposition~\ref{prop_V_hat} implies that $\hat W(r)$ is a lower bound
of $W(r)$ and that it corresponds to the length of the shortest path
from $s$ to $r$ on the extended directed graph $\tilde G$, with
arc-length given in accordance to~\eqref{eqn_bound}, namely by the
value of function $\hat \eta$.
Hence, $\hat W(f)$ can be computed by Dijkstra's
algorithm (which is equivalent to compute $V$ with A$^*$ algorithm,
with heuristic $h$). The algorithm that we are going to present is based on the
following basic observation. If A$^*$ algorithm computes
$f^*=\argmin_{f \in \hat F} \hat W(f)$ by visiting only nodes for which $\ell^+(r) \leq
\ell^-(r)$, then ii) of Proposition~\ref{prop_V_hat} is satisfied for
$r=f^*$ and $\hat W(f^*)=W(f^*)$ is the optimal value of $k$-BASP. If this is not the
case, we increase $k$ by $1$ and re-run the A$^*$ algorithm.
Note that the algorithm starts with $k=2$, since, according to its definition, $K(\BBb)=1$ only if
no acceleration bounds are present and, in this case, BASP is
equivalent to a standard SP and can be solved by Dijkstra's algorithm.

\begin{lgrthm}[Adaptive A$^*$ algorithm for k-BASP]
\label{alg_enh_ad}
\hfill
\begin{itemize}
\item[1)]  Set $k=2$.
\item[2)] Execute A$^*$ algorithm and,
at every visit of a new node $r$, if none of the two conditions $\ell^+(r) \leq
\ell^-(r)$ and $|r|<k$ hold, set $k=k+1$ and repeat step 2).
\end{itemize}
\end{lgrthm}

\begin{figure}
  \begin{center}
  \begin{tikzpicture}
    \tikzstyle{every node}=[circle, draw, fill=black,
                        inner sep=0pt, minimum width=4pt]
  \node [label=$s$] (s)                    {};
  \node         (1) [above right=2cm and 2cm of s,label=$1$] {};
  \node     (f) [right= 4cm of s, label=$f$] {};
  \path [->] (s) edge (1);
  \path[->] (1) edge (f);
  \path[->] (s) edge (f);

 \end{tikzpicture} 
    \caption{Simple graph considered in Example~\ref{example_1}.}
    \label{fig:example_1}
\end{center}
\end{figure}
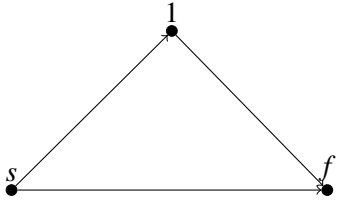

Note that the algorithm does not compute the exact value
$K(\BBb)$. Rather, it underestimates it. More precisely, it stops with the
smallest $k$ value needed to solve BASP problem between the given
source and destination nodes. This is illustrated by the following
example.

\begin{xmpl}
  \label{example_1}
Let
\[
\Gbb=(\Vbb,\Ebb),\ \ \ \Vbb=\{s,1,f\},\ \ \ \Ebb=\{(s,1), (1,f), (s,f)\},
\]
be the graph represented in Figure~\ref{fig:example_1},
with the following set of bounds $\BBb$:
\[
\begin{array}{lll}
(s,1) & \rightarrow & \alpha^-=-1 ,\ \alpha^+=1 ,\ \mu^-=0,\ \mu^+=4 \\ [6pt]
(1,f) & \rightarrow & \alpha^-=-1,\ \alpha^+=1 ,\ \mu^-=0,\ \mu^+=4 \\ [6pt]
(s,f) & \rightarrow & \alpha^-=-2 ,\ \alpha^+=2 ,\ \mu^-=0,\ \mu^+=3,  
\end{array}
\]
and edge lengths 
\[
\ell((s,1))=2,\quad \ell((1,f))=2,\quad \ell((s,f))=3.
\]
The speed is further bounded to be equal to 0 both in $s$ and in $f$.
In this case it is easily seen that $K(\BBb)=3$, since along path
$s1f$ the maximum speed is never reached under the given bounds on the
acceleration and the graph does not contain paths with more vertices.
However, the $A^*$ algorithm is first run with $k=2$. With such value,
the heuristic has the following value for the different paths of
length less or equal than $k=2$
\begin{align*}
h(s)=1,\quad h(1)=0.5,\quad h(f)=0,\\
h(s1)=0.5,\quad h(1f)=0,\quad h(sf)=0.
\end{align*}
These are easily computed by solving an SP problem with edge lengths equal to
\[
d_{s1}=0.5,\ \ d_{1f}=0.5,\ \ d_{sf}=1,
\]
obtained by the formula $d_e=\frac{\ell(e)}{\mu^+(e)}$ for each edge $e$.
The queue ${\cal Q}$ is then initialized with $\{(s,1)\}$ with $\textproc{value}(s)=0$.
Next, we remove $(s,1)$ from the queue and set $\textproc{closed}(s)=1$, and we insert in the queue 
$(1,h(s)+T_\BBb(s1)-T_\BBb(s)+h(1)-h(s))=(1,2.5)$ and $(1,h(s)+T_\BBb(sf)-T_\BBb(s)+h(f)-h(s))=\left(1,\sqrt{6}\right)$, and we set 
\[
\textproc{parent}(1) = \textproc{parent}(f) = s.
\]
Thus,
\[
{\cal Q}=\left\{\left(f,\sqrt{6}\right), (1,2.5)\right\}.
\]
Since the minimum is attained by $\left(f,\sqrt{6}\right)$, we remove it from the queue, we check whether $\ell^+(sf)\leq \ell^-(sf)$, which is the case since  $\ell^+(sf)=\ell^-(sf)=1.5$, and we stop since we reached the target node $f$. 
The minimum path is recovered from $\textproc{parent}$ (in this case it is simply path $sf$) and the minimum time to travel from $s$ to $f$ is $\sqrt{6}$.
\end{xmpl}

\begin{prpstn}
Algorithm~\ref{alg_enh} terminates with $k \leq K(\BBb)$ and
returns an optimal solution.
\end{prpstn}

\begin{proof}
By Definition of $K$, if $k=K(\BBb)$ condition $\ell^+(r)\leq\ell^-(r)$
is satisfied for all $r$. Hence, there exists $k \leq K(\BBb)$ for which the
algorithm terminates.
Let $r \in \Vbb_k$, with $\vec{r} \in F$ be the
last-visited node before the termination of the algorithm.
By ii) of Proposition~\ref{prop_V_hat}, we have that $\hat
W(r)=W(r)=V(r)$ (since $h(r)=0$),
but, by definition, $V(r)$ is the shortest time for reaching a node in
$F$.
\end{proof}

\section{Numerical experiments}
\subsection{Nodes associated to different orientations}
\label{sec_orient_graph}
\begin{figure}[!tbh]
    \centering
    \includegraphics[width=0.85\columnwidth, trim=0 85 0 53, clip=true]{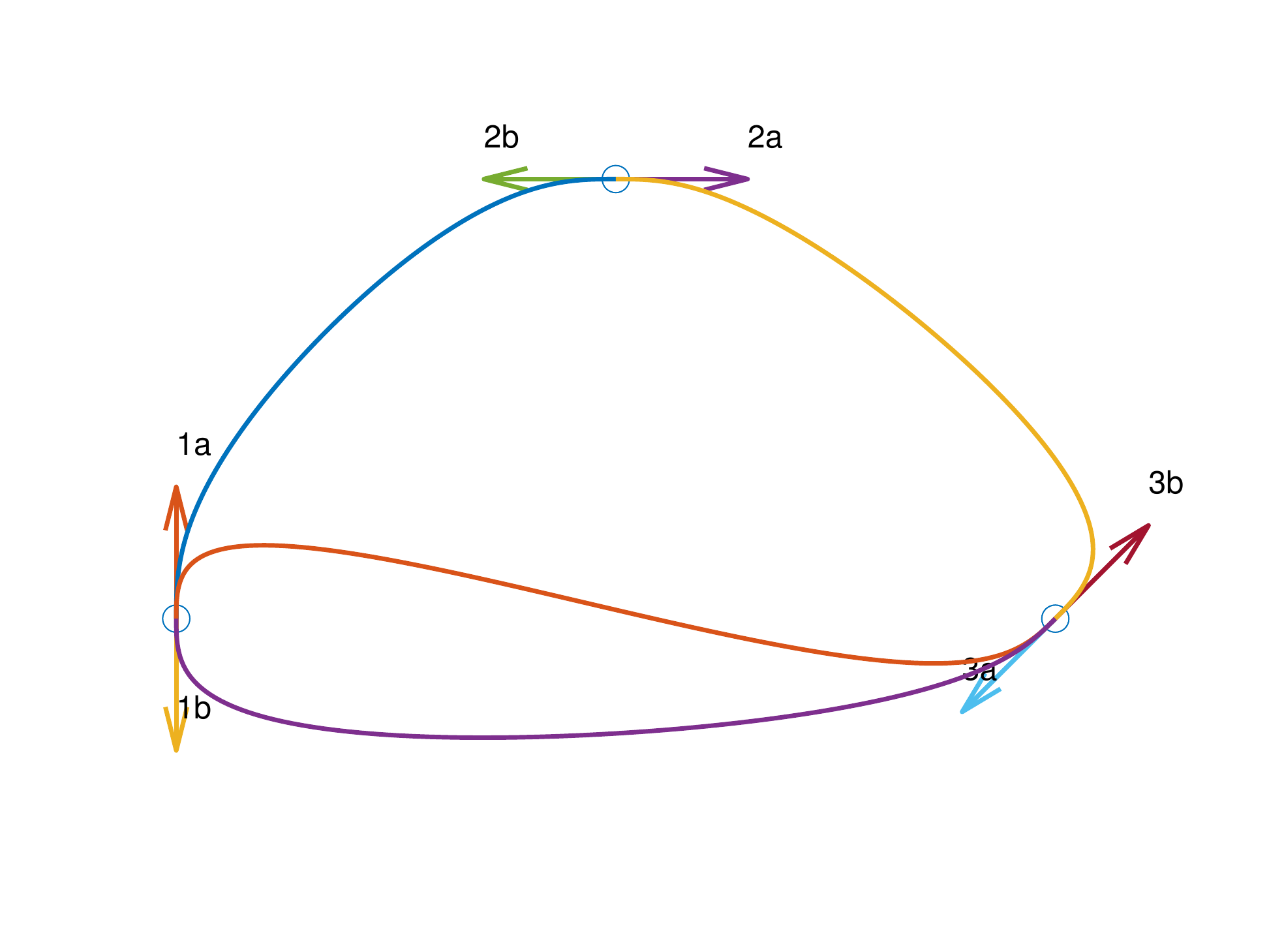}
    \caption{Graph with replicated nodes for the two possible directions.}
    \label{fig:graph_bis}
\end{figure}

\begin{figure}
\centering
\digraph[scale=0.5]{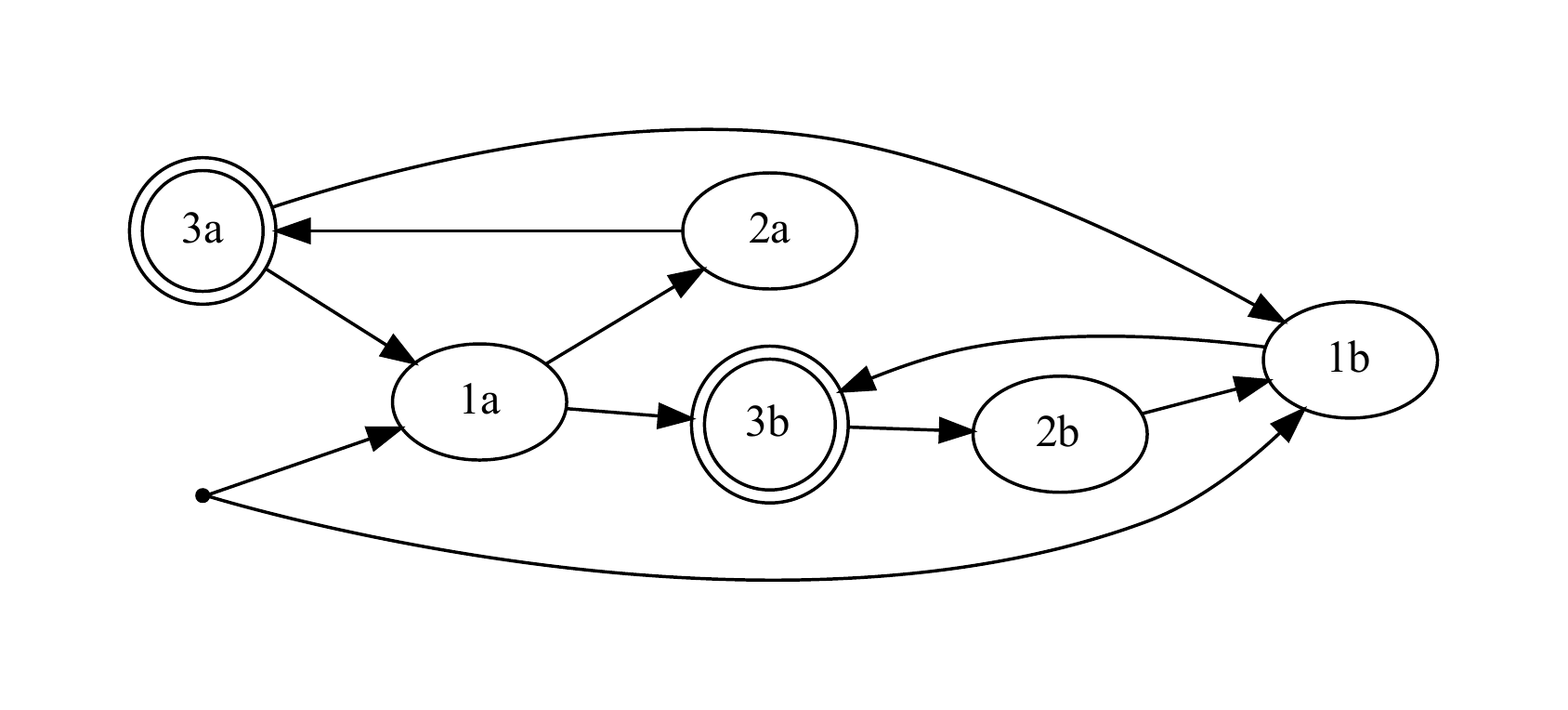}{
  rankdir=LR;
  "3a" [shape = doublecircle];
  "3b" [shape = doublecircle];
  "1a"->"2a";
  "2b"->"1b";
  "2a"->"3a";
  "3b"->"2b";
  "3a"->"1a";
  "1b"->"3b";
  "1a"->"3b";
  "3a"->"1b";
    init [label="", shape=point];
    init-> "1a";
    init->"1b";
}
\caption{Directed graph associated to the setting in Figure~\ref{fig:graph_bis}.}
\label{fig:graph_ass}
\end{figure}

Consider the setting represented in Figure~\ref{fig:graph_bis}. There
are $3$ positions connected by $4$ paths. The paths are given by
spline curves and are chosen in order to have a nonzero continuous
first derivative at connection
points. In this way, the path obtained by combining two adjacent arcs
has piecewise-continuous curvature. In order to
associate a graph to the setting of Figure~\ref{fig:graph_bis}, we
actually need to assign two nodes to each position, associated to
opposite curve directions. For instance, there is a direct arc from
node $1a$ to node
$2a$, but not from $1a$ to $2b$, since node $2b$ is associated to a
direction which is opposite to the one that we would obtain by
following the path from the first to the second position.
In this way, the setting of Figure~\ref{fig:graph_bis} is associated
to the graph reported in Figure~\ref{fig:graph_ass}
Here, position $1$ is the initial one and is associated to the two
initial nodes $1a$ and $1b$. This is due to the fact that we assume
that the vehicle is initially at rest, so that it can start
along both directions associated to $1a$ and $1b$. Similarly, final position $3$ is associated to the two states $3a$, $3b$, due to
the fact that we accept both orientations for reaching the final
position.
Handling two initial states is not problematic, since it is sufficient
to solve the problem twice, starting from both initial states $1a$
and $1b$, and then choosing the best solution.

\subsection{A 16-vertex graph}

\begin{figure}[!tbh]
    \centering
    \includegraphics[width=0.85\columnwidth, trim=0 25 0 25, clip=true]{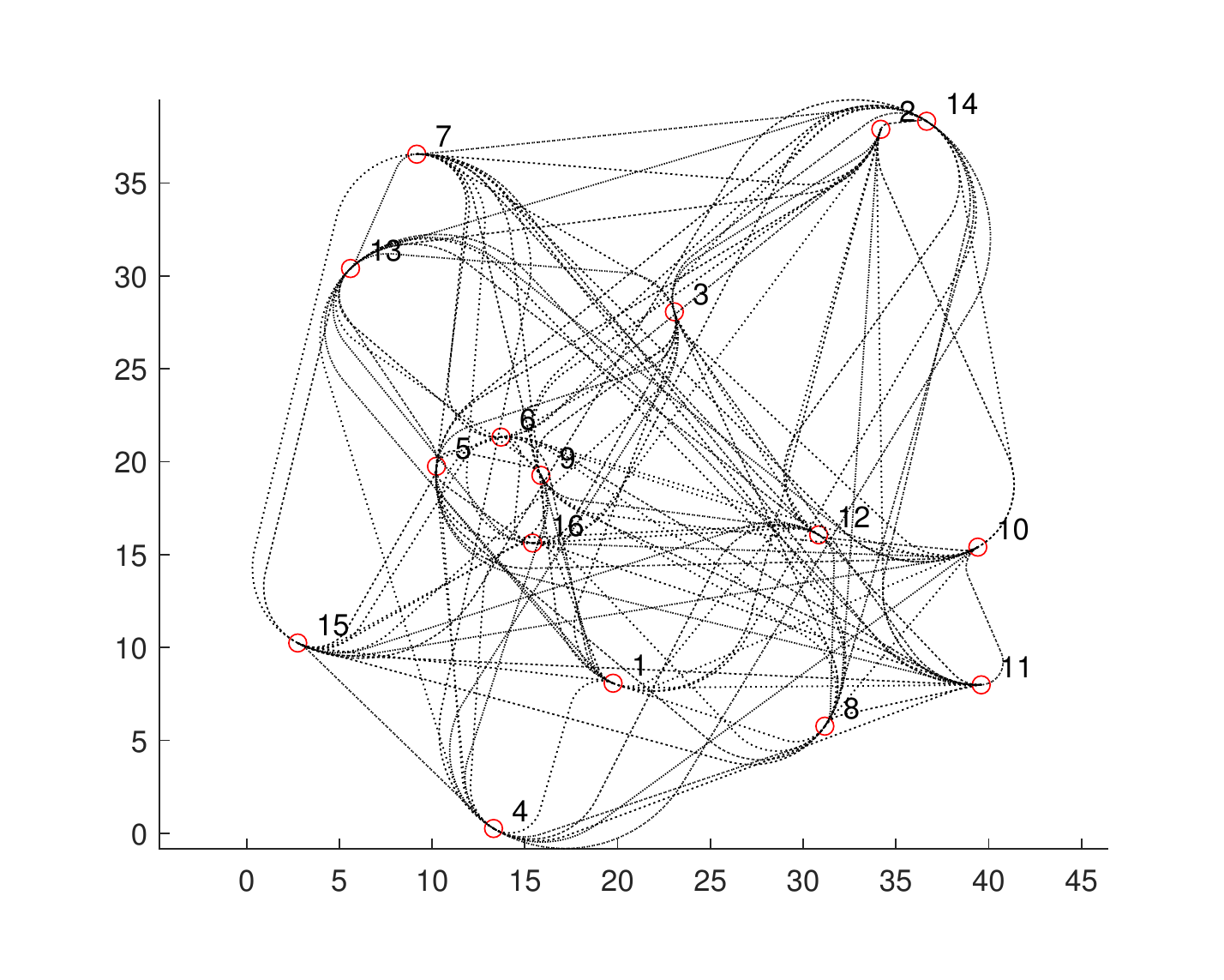}
    \caption{Graph with 16 nodes.}
    \label{fig:ex}
\end{figure}

We run all simulations on an Intel core i5 (7200u) with 8GB of RAM.
As a simple example, we consider the 16-configuration setting represented in
Figure~\ref{fig:ex}.
Each configuration $i \in \{1,\ldots,16\}$ is associated to a direction $\theta_i \in
[0,2\pi]$ and to a position $Q_i \in \Real^2$.
According to the method presented in Section~\ref{sec_orient_graph},
we associate the setting of Figure~\ref{fig:ex} to a graph with $32$
nodes.
In order to satisfy the maximum acceleration constraint, for each edge
we set the constant squared speed bound
$\mu^*((i,j)) = a_N r_{ij}$, where $r_{ij}$ is the minimum
curvature radius of the path that connects $Q_i$ to $Q_j$.
The normal acceleration $a_N= 2$ $\si{\metre\per\second^2}$ and the
maximum tangential acceleration and deceleration $\alpha^+ =- \alpha^-
= 0.5$ $\si{\metre\per\second^2}$ are constant and equal for all arcs.



\begin{figure}[!tbh]
    \centering
    \includegraphics[width=1\columnwidth, trim=0 25 0 15, clip=true]{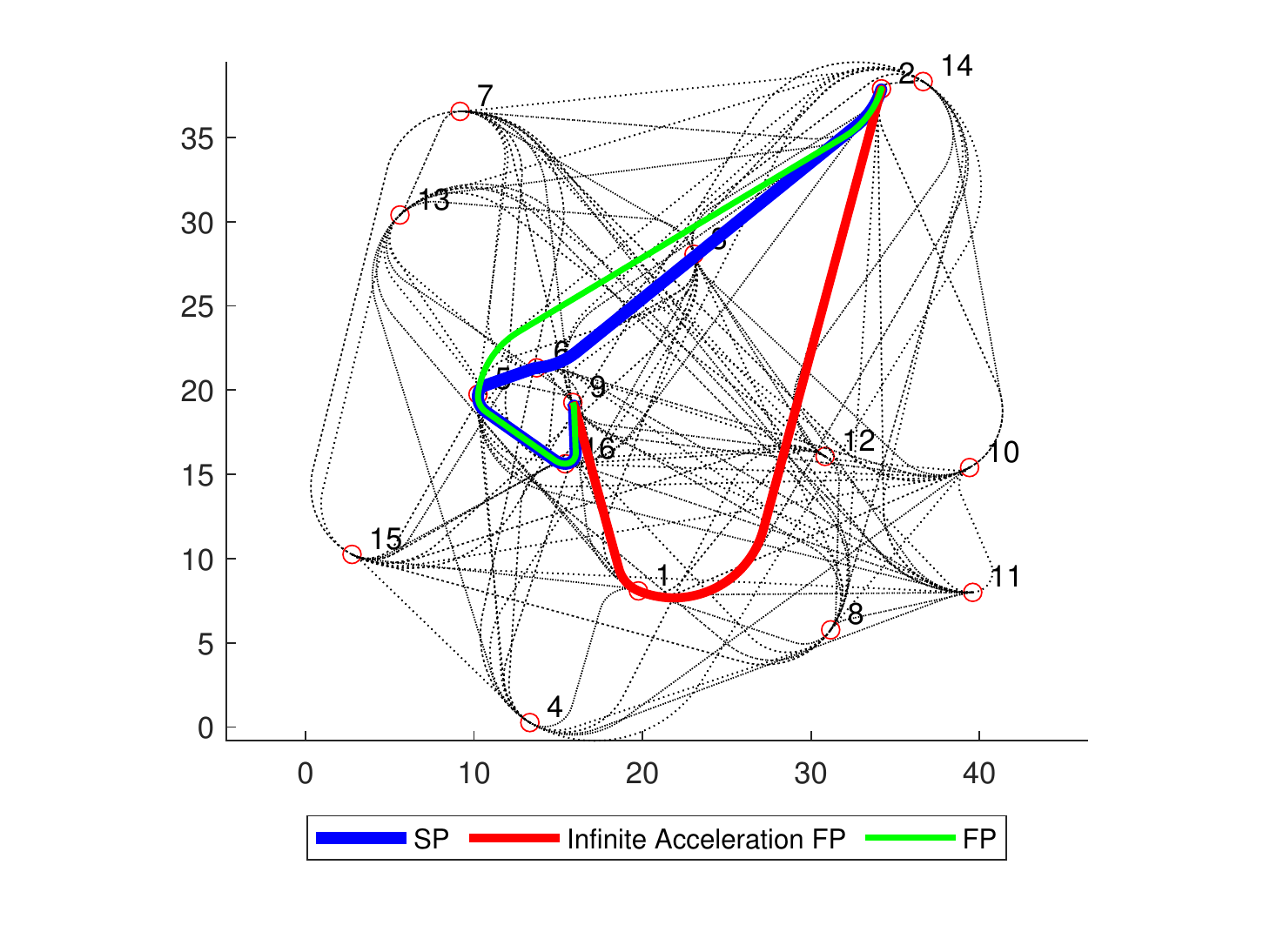}
    \caption{The three different solutions of BASP, $1$-BASP and SP.}
    \label{fig:3solutions}
\end{figure}

\begin{figure}[!tbh]
    \centering
    \includegraphics[width=0.9\columnwidth, trim=0 0 0 15, clip=true]{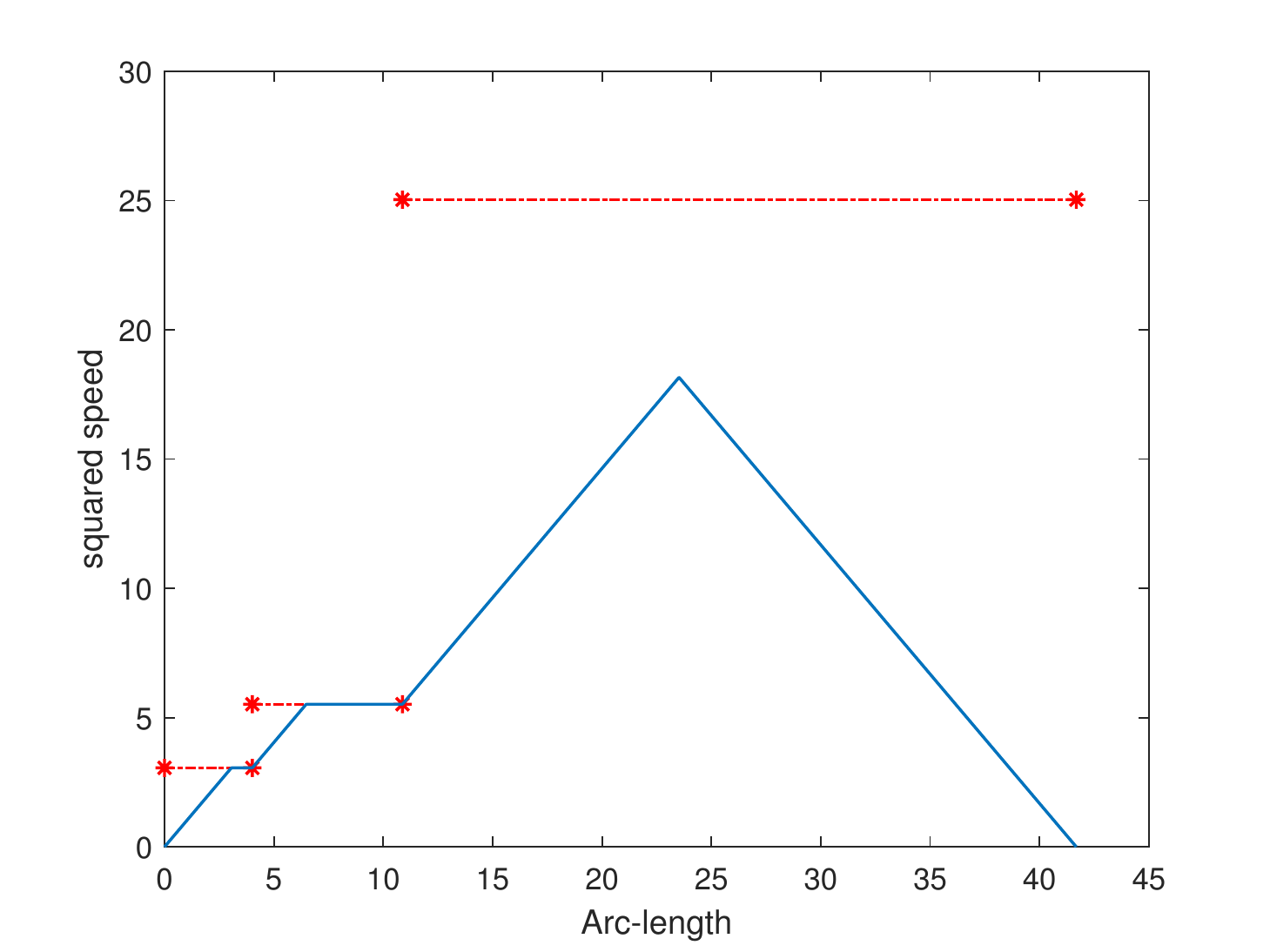}
    \caption{Speed profile of the solution of BASP with $\alpha^+ =- \alpha^- = 0.5$ $\si{\metre\per\second^2}$.}
    \label{fig:fp}
\end{figure}

\begin{figure}[!tbh]
    \centering
    \includegraphics[width=0.9\columnwidth, trim=0 0 0 15, clip=true]{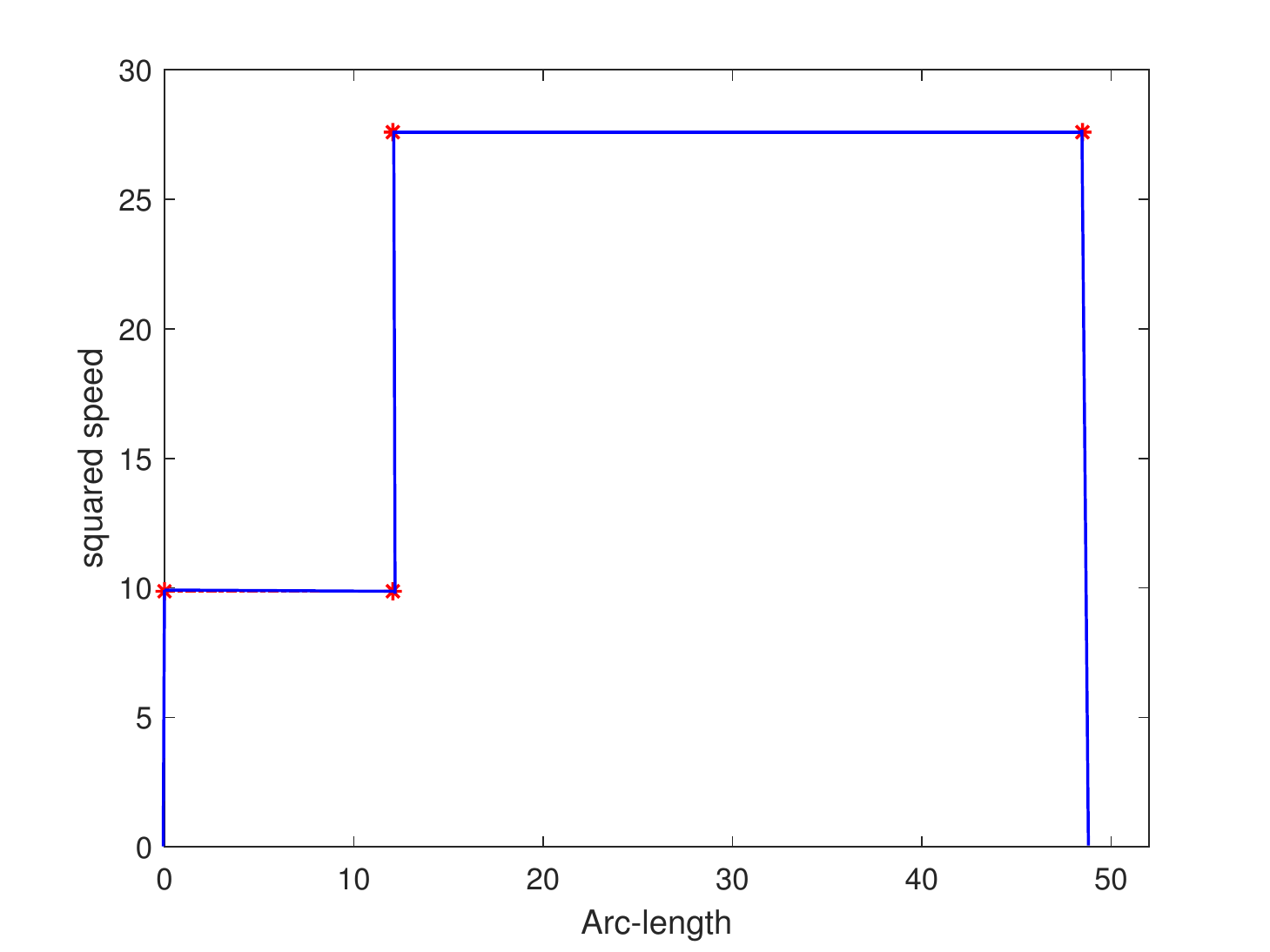}
    \caption{Speed profile of the solution of 1-BASP with infinite acceleration.}
    \label{fig:1basp}
\end{figure}

As an example, we chose as source configuration $s = 16$ and, as final one, $f = 6$,
and we computed three different solutions:
\begin{itemize}
\item  the solution of BASP;
\item the solution of BASP with infinite acceleration and deceleration
($1$-BASP);
\item the shortest path (SP).
\end{itemize}
Note that the solutions of SP and $1$-BASP can be computed by
Dijkstra's algorithm.
To compute the solution of BASP, we used Algorithm~\ref{alg_enh_ad}.
Figure~\ref{fig:3solutions} represents the solutions of the three problems.
Note that, in this case, they are all different.
In particular, in Figure~\ref{fig:fp}, we show the speed profile of the solution
of BASP, while, in Figure~\ref{fig:1basp}, we
show the speed profile of the solution of $1$-BASP, which is the
solution of BASP with infinite acceleration.
Observe that the path obtained as the solution of $1$-BASP, being $49$
\si{\metre} long, is longer than the path obtained as the solution of
BASP, which is $42$ \si{\metre} long. However, if we allow infinite
acceleration, this longer path is the minimum-time one.

The path corresponding to the solution of BASP changes according to
the chosen acceleration bounds.
In particular, if we choose a small enough acceleration bound, for
example $\alpha^+ =- \alpha^- = 0.1$ \si{\metre\per\second^2},  then
the path corresponding to the solution of BASP coincides with the shortest one.
Instead, if the acceleration bounds are large enough, for example $\alpha^+
= - \alpha^- =1$ \si{\metre\per\second^2}, the path corresponding to
the solution of BASP coincides with
the one obtained from the solution of $1$-BASP (i.e., the infinite acceleration fastest path).

\subsection{Randomly generated problems}
\label{sec_ran_graph}
We performed various tests on randomly generated problems of different
sizes, obtained with the following procedure.
First, we generated a random graph with $n$ nodes with Python package NetworkX
(\url{networkx.org}), using function\\
{\verb!geographical_threshold_graph!}.
Essentially,  each node is associated to a
position, obtained by choosing a random element of set
$[0,1]\times[0,1]$. The edges are randomly determined in
such a way that closer nodes have a higher connection probability. We multiplied the obtained position by factor $10 \sqrt{n}$, in order to obtain the same average nodes
density independently on $n$. 
For a more detailed description of
{\verb!geographical_threshold_graph!},
we refer the reader to NetworkX
documentation.
Then, we associated a random angle $\theta_i$ to each node, obtained from a uniform
distribution in $[0,2 \pi]$. In this way, each node of the random graph
is associated to a vehicle configuration, consisting of a position and
an angle. Set $\tau(\theta_i)=[\cos
\theta_i,\sin \theta_i]^T$. Each edge $(i,j)$ is associated to a \textit{Dubins path}, which is defined as the shortest
curve of bounded curvature that connects the configurations associated to nodes $i$ and
$j$, with initial tangent parallel to $\tau (\theta_i)$ and final
tangent parallel to $\tau (\theta_j)$. We chose the minimum turning radius for
the path associated to edge $(i,j)$ as $r_{ij}=\min\left\{ \ell((i,j))/(d(\theta_{i},\theta_j)),
4 \right\}$ where $d(x,y)$ is the angular distance between angles $x$ and $y$.

\begin{figure}[!tbh]
    \centering
    \includegraphics[width=1\columnwidth, trim=0 0 0 15, clip=true]{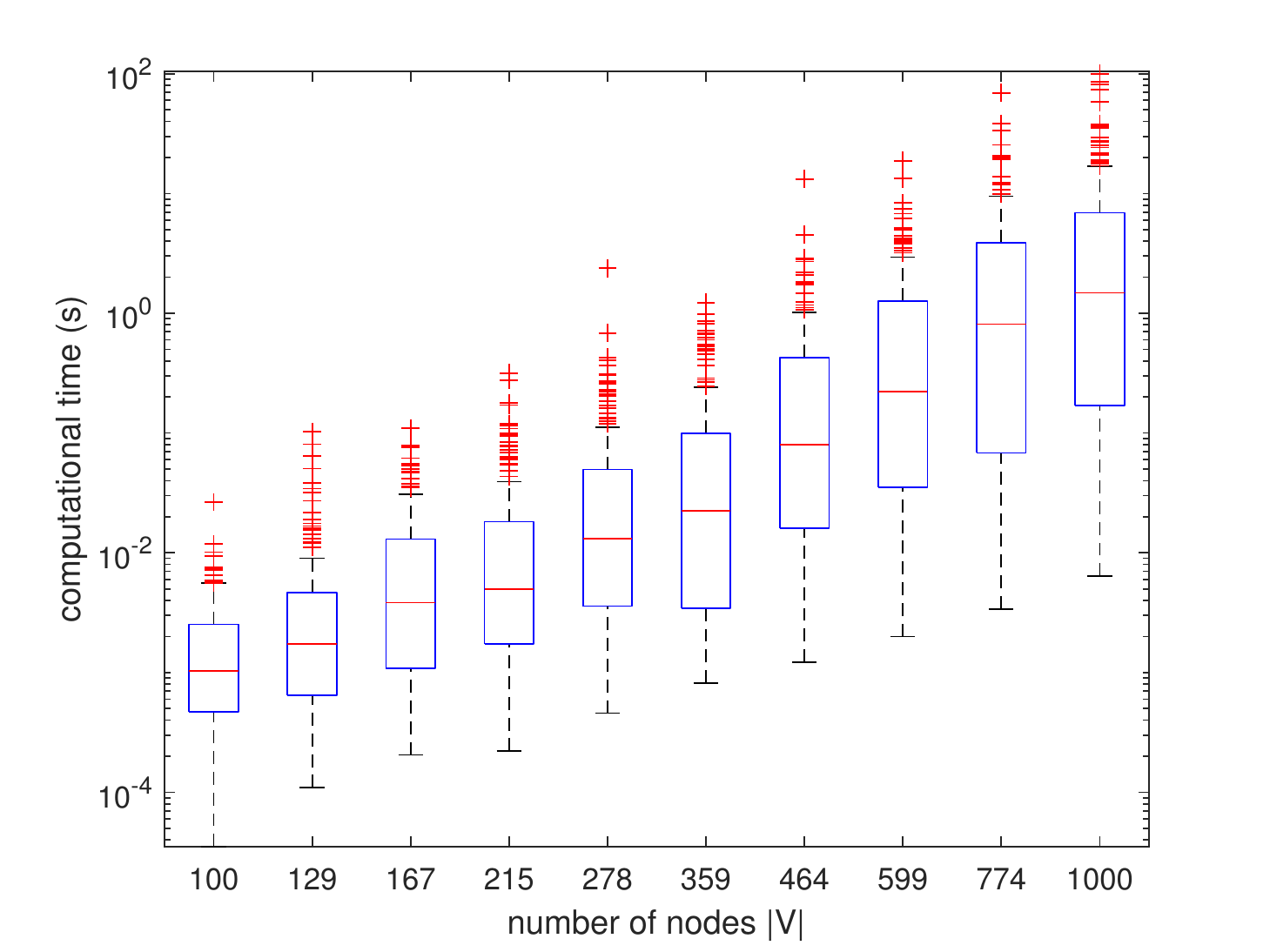}
    \caption{Box-and-whisker plot of the computational time to solve BASP
    on the different instances.}
    \label{fig:comput_time}
\end{figure}

We defined the problem graph $\Gbb$ as described in
Section~\ref{sec_orient_graph}. In particular, we associated two
nodes to each configuration, representing opposite directions. In this
way, we obtain a problem graph with $2n$ nodes.
We set the acceleration and deceleration bounds constant for all paths
and equal to $0.1$ \si{\metre\per\second^2}. The upper squared speed
bound is constant for each arc and given by $2 r$, where $r$ is the minimum
curvature radius of the path associated to the arc.
In our tests we used the adaptive A$^*$ algorithm
(Algorithm~\ref{alg_enh_ad}). First, we
ran simulations for 10 values of $n$, logarithmically spaced between $100$
and $1000$.
For each value of $n$, we generated 20 different graphs and, for each
one of them, we ran 10 simulations, randomly choosing the source and the
target node.
Figure~\ref{fig:comput_time} shows the mean values and the distributions of the computational time.
We considered as solved only those instances for which the algorithm took less than
$100$ seconds to find the solution: for $n=1000$, $5\%$ of the
instances have not been solved, while all the instances
have been solved for the other values of $n$.
Table~\ref{tab:k} shows, for each value of $n$, the percentages of
the tests in which Algorithm~\ref{alg_enh_ad} terminates with a given
value of $k$. 
 
\begin{table}[h!]
\centering
\begin{tabular}{c|c|c|c|c}
number of nodes & $k=3$  & $k=4$ &  $k=5$ &  $k=6$ \\ \hline
100 &  90 \% & 10\% & - & - \\
129 & 81.5 \% &  17.5\% &   - & - \\
167 & 70.5\% & 28\% & 1\% & 0.5\% \\
215 &  73.5\% & 25\% &  1.5 \% & - \\
278 &  64.5\% & 30.5\% & 5 \% & - \\
359 & 67.5 \% & 31 \% & 1.5\% & -\\ 
464 & 49 \% & 44.5\% & 6.5 \% & - \\
599  & 40 \% & 55\% & 5 \% & -\\
744  & 37.5\% & 53.5\% & 9 \% & -\\
1000 & 37.5\% & 57.9\% & 4.1\% & 0.5\%
\end{tabular}
\caption{Percentages of the values of $k$ for each dimension of the graph.}
\label{tab:k}
\end{table}




In the previous section, we showed that, for a given problem instance,
the path $p^*$ corresponding to the
solution of BASP is in general different from the path $\hat p$ obtained
as the solution of BASP with infinite
acceleration bounds. We ran some numerical experiments to compare the
travel times $T_{\BBb}(p^*)$ and $T_{\BBb}(\hat p)$.
Namely, we generated 50 different random graphs with $n=100$ with the
procedure presented above. For each instance, we considered $10$
problems obtained by randomly choosing the source node and the target
node. Then, we solved BASP with different acceleration bounds. Namely,
for each problem instance, we considered
equal and
constant maximum acceleration and deceleration bounds, chosen in the
range $[0.1,5.6]$ \si{\metre\per\second^2}.

\begin{figure}[!tbh]
    \centering
    \includegraphics[width=1\columnwidth, trim=0 0 0 15, clip=true]{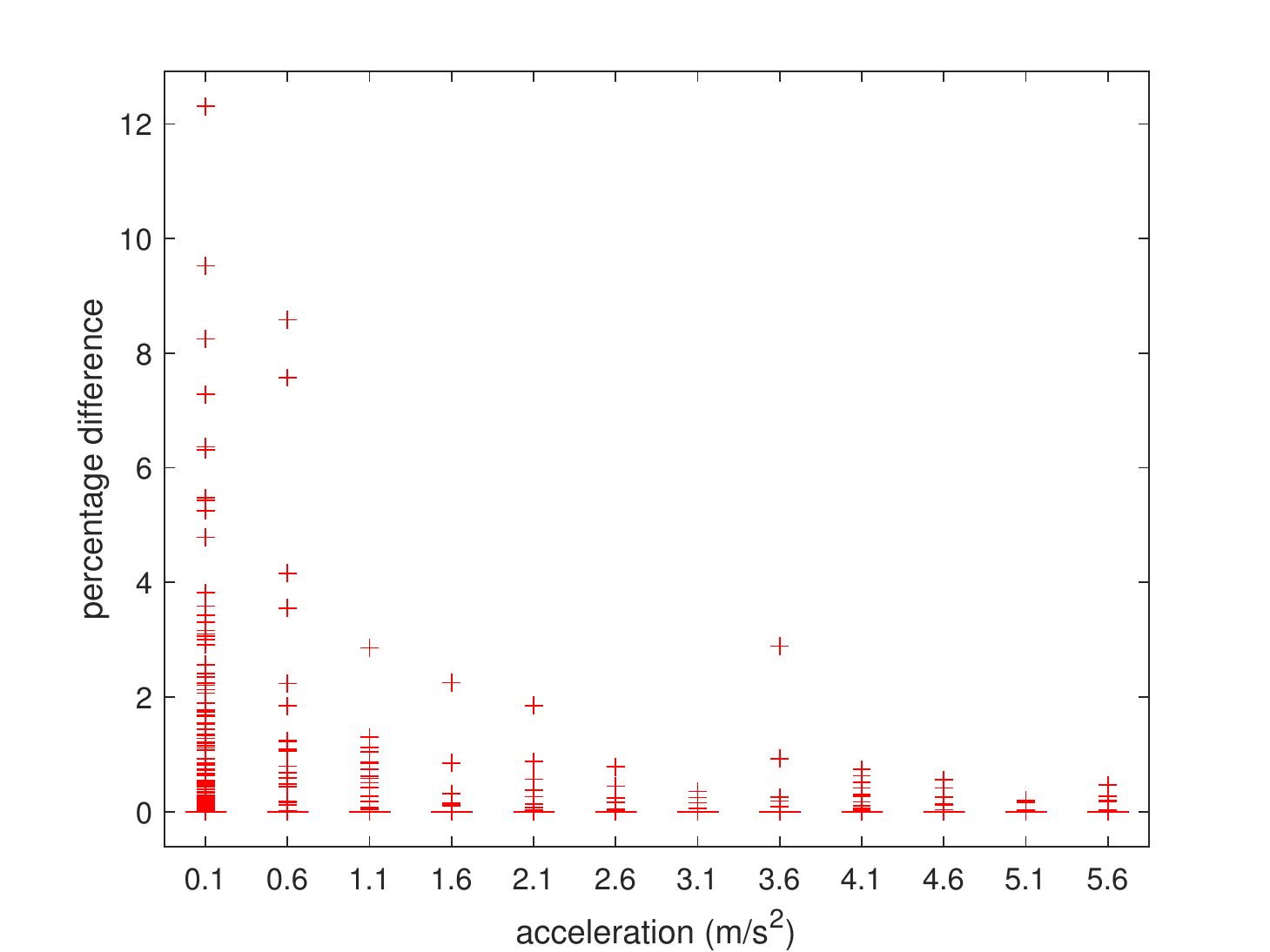}
    \caption{Percentage difference  between the travel time of the infinite acceleration FP and the travel time of the FP.}
    \label{fig:difference}
\end{figure}

In Figure~\ref{fig:difference}, we compare the optimal travel times
along the two paths. Namely, for each value of the acceleration and
deceleration bounds, we report the percentage difference
$100 \frac{T_{\BBb}(\hat p)-T_{\BBb}(p^*)}{T_{\BBb}(p^*)}$ obtained for
each test.

We observe that for low acceleration and deceleration bounds the
difference is significant, while as the acceleration and deceleration
bounds increase, the travel time difference between the two paths
tends to be smaller.
This is due to the fact that, if the acceleration/deceleration bounds
are sufficiently high, paths $p^*$ and $\hat p$ are the same.

\subsection{Real industrial applications}
Here we present two problems taken from real industrial applications,
representing
two automated warehouses.
The problem data have been provided by packaging company Ocme S.r.l.,
based in Parma, Italy.
The first problem is described by a graph of 399 nodes. The
acceleration and deceleration bounds are constant, equal for all
paths, and given by
$\alpha^+ = 0.28$ \si{\metre\per\second^2} and $\alpha^- = -0.19$
\si{\metre\per\second^2}. The speed bounds are constant for
each arc but vary among different arcs, according to the associated
paths curvatures, and they take values in the interval $[0.136. 1.7]$ \si{\metre\per\second}.
The arc-lengths take values between $0.628$ \si{\metre} and $10.87$ \si{\metre}
and have an average value of $2.86$ \si{\metre}.

\begin{figure}[!tbh]
    \centering
    \includegraphics[width=1\columnwidth, trim=0 20 0 12, clip=true]{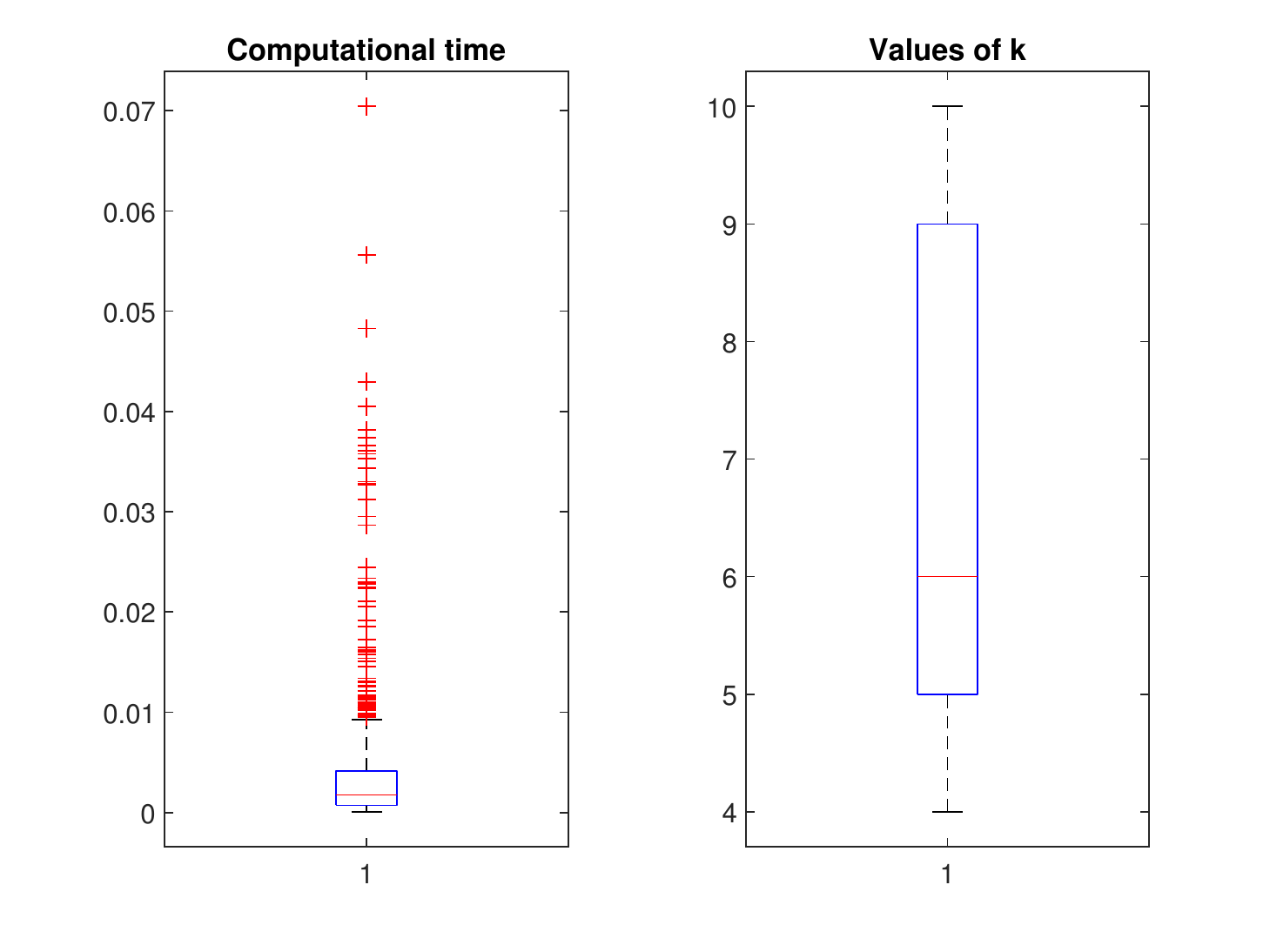}
    \caption{Box-and-whisker plot of the 1000 simulations on the 399-vertex graph.}
    \label{fig:small}
\end{figure}

We ran 1000 simulations by randomly choosing the source node and the
target node.
The average value and the standard deviation of the computational time
are $0.0036$ \si{\second} and $0.0062$ \si{\second}, respectively.
In
Figure~\ref{fig:small}, we report the distribution of the
computational time of the considered instances.
In the same figure we also show a box-and-whisker plot that reports
the final value of $k$ obtained by Algorithm~\ref{alg_enh_ad} for solving each instance.

We also considered a second problem, representing a larger automated
warehouse, described by a graph of 3399 nodes.
The acceleration and deceleration ramps are the same as in
the previous example, while the maximum speed bounds belongs to the
interval 
$[0.086, 1.7]$
\si{\metre\per\second}. The arc-lengths take values between $0.2$
\si{\metre} and $16.352$ \si{\metre} and have an average value of $3.569$ \si{\metre}. 

As in the first example, we ran 1000 simulations by randomly choosing the
source and the target nodes and we found that the average value and the
standard deviation of the computational time are $0.0128$ \si{\second} and 
$0.0058$ \si{\second}, respectively.

\begin{figure}[!tbh]
    \centering
    \includegraphics[width=1\columnwidth, trim=0 20 0 12, clip=true]{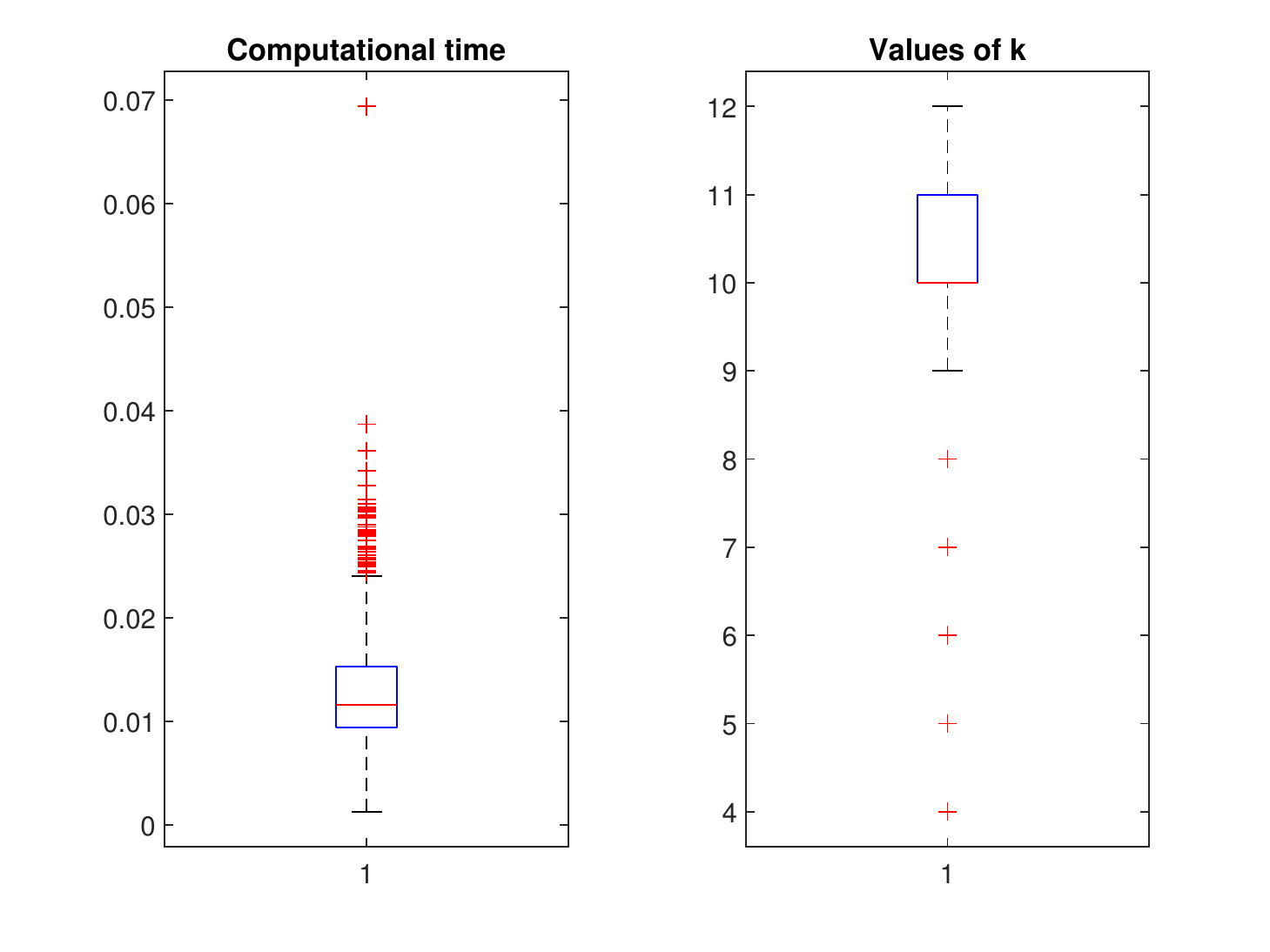}
    \caption{Box-and-whisker plot of the 1000 simulations on the 3399-vertex graph.}
    \label{fig:big}
\end{figure}

In Figure~\ref{fig:big}, we report the box-and-whisker plots of the computational
time and of the final value for $k$ in Algorithm~\ref{alg_enh_ad}, for each instance.
Note that both the mean computational time and the final value of $k$
of the second example are larger than those of the first one. This is
due to the fact that the second problem has a larger number of
nodes.  We can also note that the mean computational times of these
two real-life examples are much lower than those of the random tests
of comparable size
presented in Section~\ref{sec_ran_graph}. 
This is probably due to the fact that the graphs associated to the two
industrial problems have a lower connectivity than the randomly
generated ones. Indeed, most nodes in the two industrial
problems represent positions in corridors and are connected only to
two other nodes: the preceding and the following one along the
corridor. Note that this is common in problems associated to automated
warehouses, since these facilities often have many long corridors.

\subsection{Example with non constant speed bounds}
\begin{figure}[!tbh]
	\centering
	\includegraphics[width=1\columnwidth, trim=0 18 0 20, clip=true]{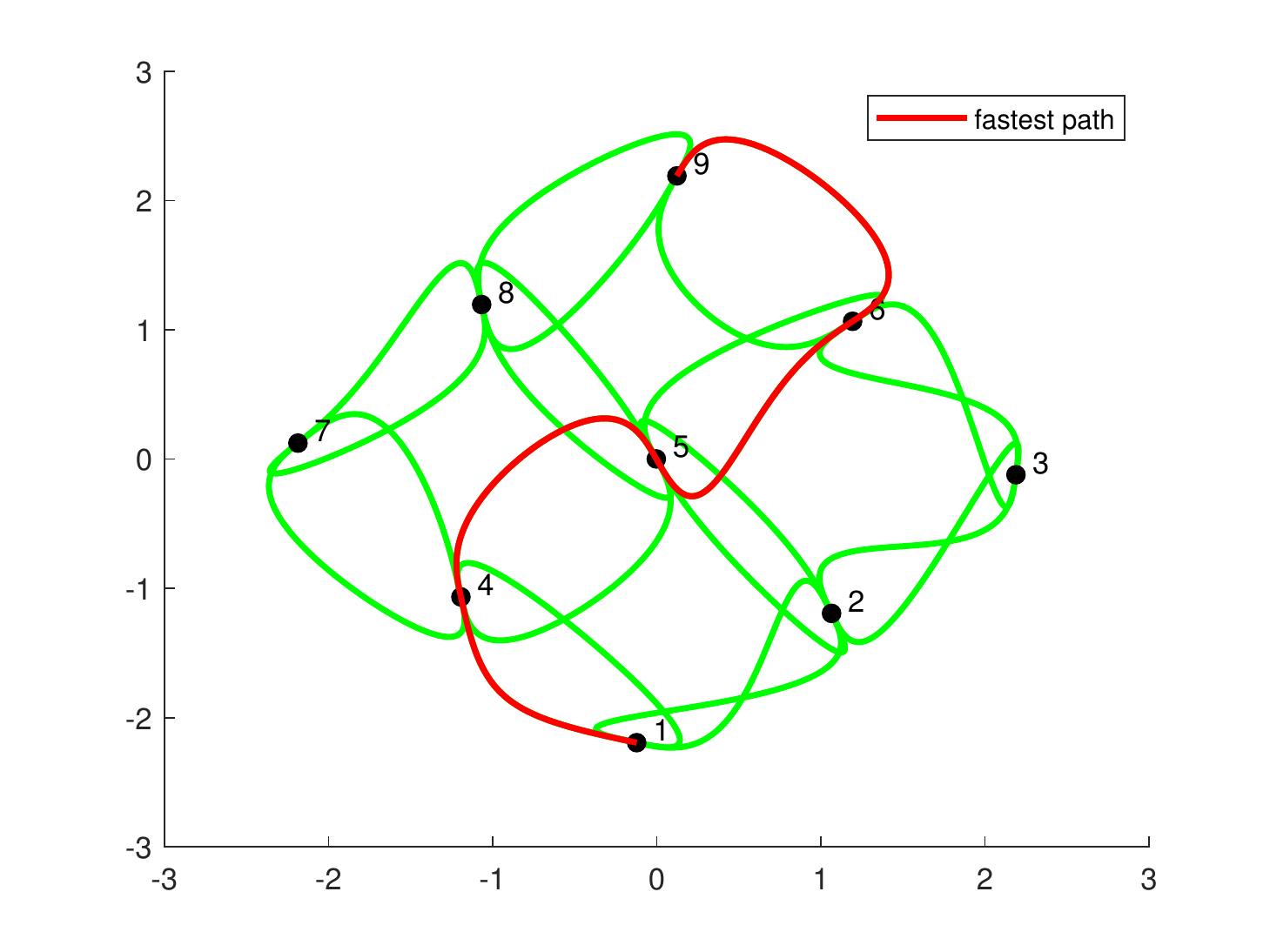}
	\caption{Graph with 9 nodes.}
	\label{fig:9graph}
\end{figure}

In all previous simulations, we considered problem instances in which
acceleration and speed bounds are constant along each arc.
However, the setting of BASP, as defined in~\ref{prob_BASP}, allows
for arc-length dependent bounds on each arc.
Here, we considered a problem instance of this more general form,
illustrated by Figure~\ref{fig:9graph}.
We considered $9$ configurations, each one associated to a position
on the plane and to a direction angle. We defined the connecting
paths by an order 5 interpolating spline, with initial and final
conditions that guarantee the continuity of the
tangent vector on connection points.

We choose a maximum speed bound $v_{\max} = 1.7$ \si{\metre\per\second} and a maximum 
normal acceleration $a_N = 0.56$ \si{\metre\per\second^2}.
The speed bound is a continuous function defined at each point $\lambda$
of a path as $v(\lambda)= \min\left\{v_{\max},\sqrt{a_N/|\kappa(\lambda)|} \right\}$,
where $\kappa$ is the scalar curvature of the path, which is a function whose
absolute value is the inverse of the radius of the circle that locally
approximates the geometric path.

\begin{figure}[!tbh]
	\centering
	\includegraphics[width=1\columnwidth, trim=0 27 0 15, clip=true]{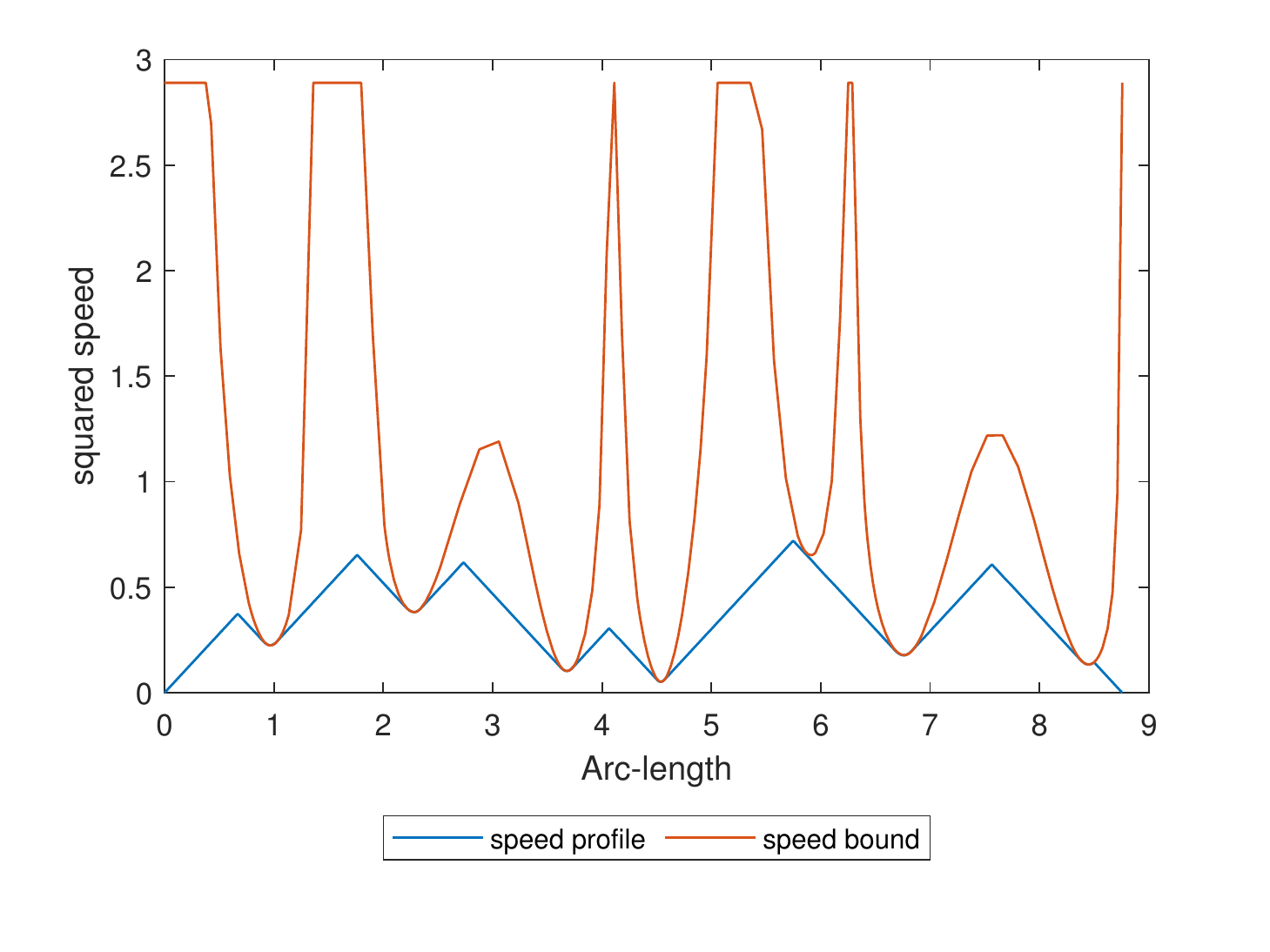}
	\caption{Speed profile of the fastest path.}
	\label{fig:9_graph_speedprofile}
\end{figure}

Figure~\ref{fig:9graph} also shows the solution of BASP, with source
node $s = 1$, while Figure~\ref{fig:9_graph_speedprofile} shows the corresponding speed profile.


\section{Conclusions}
The main contributions of this work are the definition of BASP, the
proof of its NP-hardness, and the definition of a solution
algorithm that achieves polynomial time-complexity under some hypotheses on problem data.



\section*{Appendix}
\begin{prpstn}
  \label{prop_F1_F2}
Let $\mu,\alpha:[0,+\infty) \to \Real^+$, let $F_1,F_2$ be the
solutions of the following equations,
\begin{equation}\label{eqn_forf}
\begin{cases}
F_i'(\lambda) =
\begin{cases}
\alpha(\lambda)	\wedge \mu'(\lambda)	& \text{if } F_i(\lambda) \geq \mu(\lambda) \\
\alpha(\lambda)						& \text{if } F_i(\lambda) < \mu(\lambda)
\end{cases}\\
F_i(0) =  w_{0,i},
\end{cases}
\end{equation}
with $0 \leq w_{0,i} \leq \mu(0)$, for $i \in \{1, 2\}$, and let $\bar \lambda$
be such that $\mu(\bar \lambda)=\int_0^{\bar \lambda} \alpha(\lambda) d\lambda$.
Then $(\forall \lambda \geq \bar \lambda)\ F_1(\lambda)=F_2(\lambda)$.
\end{prpstn}

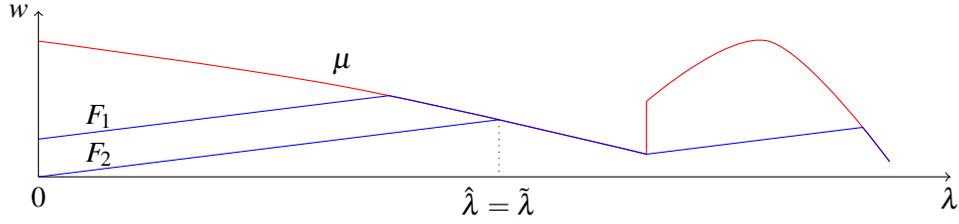
\begin{figure}
\begin{center}
\begin{tikzpicture}[y=1cm, x=8cm]
\draw[->] (0,0) -- (1.5,0) node[anchor=north] {$\lambda$};
\draw[->] (0,0) -- (0,2.2) node[anchor=east] {$w$};
\draw [name path=primo, red] plot [smooth, tension=0.6] coordinates { (0,1.8) (0.5,1.2)
(1,0.3)};
\draw [name path=parte2, red] plot [smooth, tension=0.6] coordinates
{(1,1)(1.2,1.8)(1.4,0.2)};
\draw [red] (1,0.3)--(1,1);
\draw [name path=secondo, draw=none](0,0)--(1,1);
\path [name intersections={of=primo and secondo,by=inter}];
\draw [blue] (0,0)--(inter);
\draw [name path=terzo, draw=none](0,0.5)--(1,1.5);
\path [name intersections={of=primo and terzo,by=inter2}];
\draw [blue] (0,0.5)--(inter2);
\draw [name path=quarto, draw=none](1,0.3)--(2,1.3);
\path [name intersections={of=parte2 and quarto,by=inter3}];
\draw [blue] (1,0.3)--(inter3);
\coordinate (orig) at (0,0){};
\draw [dotted] (inter)--(inter|-orig); 
\draw (0.1,0.3) node {$F_2$};
\draw (0.1,0.8) node {$F_1$};
\draw (0.5,1.5) node {$\mu$};
\draw (0,0) node[anchor=north] {$0$}
 (inter|-orig) node[anchor=north] {$\hat \lambda=\tilde \lambda$}; 
 
\begin{scope}
\clip(inter3) rectangle ++(1,-1);
    \draw [blue] plot [smooth, tension=0.6] coordinates
{(1,1)(1.2,1.8)(1.4,0.2)};
\end{scope}

\begin{scope}
\clip(inter2) rectangle ++(1,-1);
\draw [blue] plot [smooth, tension=0.6] coordinates {(0,1.8) (0.5,1.2)
(1,0.3)};
\end{scope}
\end{tikzpicture}

\caption{Illustration of the proof of Proposition~\ref{prop_F1_F2}.}
\label{fig:proof1}
\end{center}
\end{figure}
\begin{proof}
Figure~\ref{fig:proof1} illustrates the following proof.
W.l.o.g., assume that $w_{0,1} \geq w_{0,2}$.
This implies that $(\forall \lambda \geq 0)\ F_1(\lambda) \geq F_2(\lambda)$.
Indeed, assume by contradiction that there exists $\bar \lambda$ such that
$F_1(\bar \lambda) < F_2(\bar \lambda)$, then, by continuity
of $F_1$ and $F_2$, this implies that there exists $\hat \lambda \leq
\bar \lambda$ such that
$F_1(\hat \lambda)=F_2(\hat \lambda)$, thus $(\forall \lambda \geq \hat \lambda)\
F_1(\lambda)=F_2(\lambda)$, since, for $\lambda \geq \hat \lambda$,
$F_1(\lambda)$ and $F_2(\lambda)$ solve the same differential equation
with the same initial condition at $\lambda = \hat \lambda$, contradicting the assumption.

Further, note that $(\exists \tilde{\lambda} \in (0, \bar{\lambda}])\
F_2(\tilde{\lambda}) = \mu(\tilde{\lambda})$.
Indeed, if by contradiction
\[
(\forall \lambda \in(0,\bar \lambda])\ F_2(\lambda)<\mu(\lambda),
\]
then
\[
(\forall \lambda \in(0,\bar \lambda])\ F_2'(\lambda)=\alpha(\lambda),
\]
so that
\[
F_2(\bar \lambda) - F_2(0) = \displaystyle \int_0^{\bar \lambda}
\alpha(\lambda)\; d\lambda = \mu(\bar \lambda),
\]
which contradicts the assumption.

Hence, $(\exists \hat {\lambda} \in \Real^+)\ F_2(\hat{\lambda})
=F_1(\hat \lambda)=
\mu(\hat{\lambda})$ and, consequently, 
\[(\forall \lambda \geq \hat{\lambda})\ F_1(\lambda)=F_2(\lambda),\]
which implies the thesis, being $\bar \lambda \geq \hat \lambda$.
\end{proof}

For $p \in P$, $\lambda \in [0,\ell(p)]$, we set
$\Wc_{p}(\lambda)=w$, where $w$ is the solution of
Problem~\eqref{eqn_problem_pr_wg} for path $p$. In other words
$\Wc_p(\lambda)$ is the square of the optimal speed profile for
traversing path $p$, evaluated at arc-length $\lambda$, with respect
to $p$.

\begin{prpstn}
  \label{prop_prop_W}
  
\begin{itemize}
\item[1)] Let $p_1,p_2,q \in P$, be such that $p_1q,p_2q \in P$, then
\[
(\forall \lambda \geq \ell^+ (q))\ \Wc_{p_1 q}(\ell(p_1)+\lambda)
=\Wc_{p_2 q}(\ell(p_2)+\lambda).
\]

\item[2)] Let $p,q_2,q_1 \in P$, be such that $pq_1,pq_2 \in P$, then
\[
(\forall \lambda \leq \ell^-(p))\ \Wc_{p q_1}(\lambda)=\Wc_{p q_2}(\lambda).
\]
\end{itemize}
\end{prpstn}

\begin{proof}
We only prove 1), the proof of 2) is analogous.
Note that, for $\lambda \geq 0$, $\Wc_{p_1 q}(\lambda+\ell(p_1))=
\min\{F_1(\lambda),B(\lambda)\}$, $\Wc_{p_2 q}(\lambda+\ell(p_2))=
\min\{F_2(\lambda),B(\lambda)\}$, where $F_1$, $F_2$ are the
solution of~\eqref{eqn_forf} with $\mu=\mu^+$ and initial conditions
$w_{0,1}=\Wc_{p_1}(\ell(p_1))$ and $w_{0,2}=\Wc_{p_2}(\ell(p_2))$,
respectively, and $B$ is the solution of~\eqref{B_def} with $\mu=\mu^+$.
By Proposition~\ref{prop_F1_F2}, for $\lambda \geq \ell^+(q)$,
$F_1(\lambda)=F_2(\lambda)$.
Consequently, $(\forall \lambda \geq \ell^+(q))$ $\
\Wc_{p_1q}(\ell(p_1)+\lambda)=\Wc_{p_2 q}(\ell(p_2)+\lambda)$. 
\end{proof}

\subsection{Proof of Proposition~\ref{prop_lpm}}

Let $\Psi$ be defined as in~\eqref{obj_fun_pr_wg}, then
\[\begin{gathered}
  T(p_1 t \sigma)-T(p_1 t)=\\ \int_0^{\ell(p_1 t \sigma)}
\Psi(\Wc_{p_1t \sigma}(\lambda)) d\lambda- \int_0^{\ell(p_1 t)}
\Psi(\Wc_{p_1t}(\lambda)) d\lambda =\\ \int_{\ell(p_1)+ \ell^-(t)}^{\ell(p_1 t \sigma)}
\Psi(\Wc_{p_1t \sigma}(\lambda)) d\lambda- \int_{\ell(p_1)+ \ell^-(t))}^{\ell(p_1 t)}
\Psi(\Wc_{p_1t}(\lambda)) d\lambda,
\end{gathered}
\]

where we used the fact that, by ii) of
Proposition~\ref{prop_prop_W}, $(\forall \lambda \leq \ell(p_1)+ \ell^-(t))$ $\
\Psi(\Wc_{p_1t \sigma}(\lambda))= \Psi(\Wc_{p_1t}(\lambda))$.
Similarly, we have that $T(p_2 t \sigma)-T(p_2 t)=$

 $\int_{\ell(p_2)
+ \ell^-(t)}^{\ell(p_2 t \sigma)} \Psi(\Wc_{p_2t \sigma}(\lambda)) d\lambda -
\int_{\ell(p_2) + \ell^-(t)}^{\ell(p_2 t)} \Psi(\Wc_{p_2t}(\lambda)) d\lambda$.

Moreover, by i) of Proposition~\ref{prop_prop_W}, we have that
$(\forall \lambda \geq \ell^+(t \sigma))\ \Wc_{p_1t  \sigma}(\ell(p_1)+\lambda)
d\lambda=\Wc_{p_2t \sigma}(\ell(p_2)+\lambda) d\lambda$ and
$(\forall \lambda \geq \ell^+(t))\ \Wc_{p_1t}(\ell(p_1)+\lambda) d\lambda =
\Wc_{p_2t}(\ell(p_2)+\lambda) d\lambda$ which imply that
$T(p_1 t \sigma)-T(p_1 t)=T(p_2 t \sigma)-T(p_2 t)$, since $\ell^+(t) \leq
\ell^-(t)$ and, as noticed in Section~\ref{sec_kBASP}, $\ell^+(t\sigma) \leq
\ell^+(t)$.
\qed


\subsection{Proof of Proposition~\ref{prop_NP_hard}}
\label{sec:proofNPhard}
Let $s\in \Vbb$ be the departure node and $f\in \Vbb$ be the arrival node.
Let $v_s$ be the initial speed at node $s$ and $v_f$ be the final speed
at node $f$.
We would like to select a path in $\Gbb$ from $s$ to $f$, such that the time
needed to run along the path by fulfilling the maximum speed, the
maximum and minimum acceleration constraints along the edges,
and the boundary conditions $v_s$ and $v_f$, is minimized.
We show that this problem is NP-hard by a polynomial reduction of the NP-complete {\em Partition}
problem to BASP-C.
In the Partition problem, given a set $N=\{1,\ldots,n\}$ of positive integer
values $w_1,\ldots,w_n$, we would like to establish whether $N$ can be
partitioned into two subsets $N_1$ and $N_2$ such that $\sum_{i\in N_1}
w_i=\sum_{i\in N_2} w_i=\frac{W}{2}$.
Given an instance of the Partition problem we polynomially reduce it to an
instance of BASP-C as follows.
Let $\Gbb=(\Vbb,\Ebb)$ be such that:
\[
\Vbb = N \cup \left\{0,n+1\right\}, \quad \Ebb = \{(i,j) \in \Vbb^2 \mid i<j\}.
\]
We set the following lengths for the arcs:
\[
\ell(i,j)=
\begin{cases}
0,	& i=0 \\
w_i,	& \mbox{otherwise.}
\end{cases}
\]
For what concerns the maximum speed values, we set $(\forall e \in \Ebb)\
\mu^+(e)=+\infty$ (unbounded maximum speed), while we set the
maximum acceleration $\alpha^+=1$ and the minimum acceleration
$\alpha^-=-1$ for all arcs.
The starting node $s$ is node $0$, with $v_s=0$, while the final node
$f$ is $n+1$ with $v_f=\sqrt{W}$.
Each path $p$ from node $0$ to node $n+1$ has the following structure
\[
0\ i_1\ i_2\ \cdots\ i_r\ n+1,
\]
with $0<i_1<i_2<\cdots<i_r$.
Let us denote by $N_p=\{i_1,i_2,\ldots,i_r\}$ the set of intermediate nodes in $p$.
The length of path $p$ is $\ell(p)=\sum_{i\in N_p} w_i$.
Let us first assume that $\ell(p)<\frac{W}{2}$. In this case the maximum
speed which can be reached at the end of the path is $v_u=\alpha^+ t_p$,
where $t_P$ fulfills $\ell(p)=\frac{1}{2}\alpha^+ t^2$, (i.e., $t_p=\sqrt{2 \ell(p)}$).
Thus, $v_u= \sqrt{2 \ell(p)}< v_f$, (i.e., no path $p$ with $\ell(p)<\frac{W}{2}$
is able to meet the boundary condition $v_f$).
Thus, we restrict our attention to paths $p$ such that $\ell(p)\geq \frac{W}{2}$.
A lower bound for the time needed to run along the path is given again by
the solution of the following simple equation  $\ell(p)=\frac{1}{2}\alpha^+ t^2$,
(i.e., $t_p=\sqrt{2 \ell(p)}$).
Note that this is a lower bound since, with the maximum acceleration, after this
time we reach speed $v_u=\alpha^+ t_p=\sqrt{2 \ell(p)}\geq \sqrt{W}$,
so that we might need to decelerate in order to meet the boundary condition
$v_f$.
Since $\ell(p)\geq \frac{W}{2}$, we have that the lower bound can be further
bounded from below by $\sqrt{W}$.
Finally, we observe that such lower bound can be attained if and only if the
Partition problem admits a solution.
In such case we can set $N_1=N_p$ and $N_2=N\setminus N_p$.
Thus, we have established that an instance of the Partition problem admits
a solution if and only if the corresponding instance of the BASP has
optimal value equal to $\sqrt{W}$.
\subsection{Proof of Proposition~\ref{prop_pseudo_poly}}
\label{sec:proofpseudopoly}
We modify the original problem as follows.
First, we split each arc $\theta$ into $\ell(\theta)$ arcs of length 1 by introducing
along the original arc $\ell(\theta)-1$ intermediate nodes (recall that $\ell(\theta)$ is assumed to be integer).
In this way, we have a new graph with node set $\Vbb'$ such that $|\Vbb'|
= |\Vbb|+\sum_{\theta \in \Ebb} \ell(\theta) - |\Ebb|$ and arc set $\Ebb'$ where
each arc $\theta \in \Ebb$ is replaced by $\ell(\theta)$ arcs and all arcs have
length equal to 1. The new arcs inherit the speed and acceleration bounds of the original ones.
Next, we observe that at optimal solutions there is a finite number of
speeds which can be reached at each node.
These include all squared speeds $\mu^+(\theta)$ for $\theta \in \Ebb$
but also all speeds which can be reached  starting from one squared
speed $\mu^+(\theta)$ and then moving with a maximum (or minimum)
acceleration along a path $p$ of length $\ell(p)$, provided that we never
reach the maximum speed along an arc of the path and that the speed
never falls below 0.
In order to meet the last two requirements, the value $\ell(p)$ is bounded
from above by $\frac{1}{2}(\max_{\theta \in \Ebb} \mu^+(\theta))^2$.
Indeed, the time $t_p$ required for a path $p$ of length $\ell(p)$, 
assuming that the initial speed is 0, is given by the solution of
\[
\ell(p) = \frac{1}{2}\alpha^+ t^2,
\]
so that the corresponding variation of the speed is $\alpha^+ t_p$
which needs to be lower than $\max_{\theta \in \Ebb} \mu^+(\theta)$.
Recalling that $(\forall \theta \in \Ebb)\ \alpha^+=1$, we must have that
$\sqrt{2\ell(p)} \leq \max_{\theta \in \Ebb} \mu^+(\theta)$ or, equivalently,
$\ell(p) \leq \frac{1}{2}(\max_{\theta \in \Ebb} \mu^+ (\theta))^2$.
Now, let us denote by ${\cal V}$ the set of different possible speeds.
In view of the previous observations, we have that 
$|{\cal V}|\leq |\Ebb|(1+\frac{1}{2}(\max_{\theta \in \Ebb} \mu^+ (\theta))^2)$.
Now we create a new graph with node set $\Vbb' \times {\cal V}$, (i.e.,
each node is a pair made up by a node in $\Vbb'$ and one of the possible
speeds in ${\cal V}$).
Thus, the number of nodes is 
\[
|\Vbb'| |{\cal V}| \leq (|\Vbb|+\sum_{\theta \in \Ebb}
\ell(\theta)-|\Ebb|)(|\Ebb|(1+\frac{1}{2}(\max_{\theta \in \Ebb} \mu^+ (\theta))^2)). 
\]
For what concerns the arc set, in this graph an arc between node $(i,w_i)$
and node $(j,w_j)$ exists if there exists an arc $(i,j)\in \Ebb'$.
The distance associated to this arc is the minimum time for a path from
$i$ to $j$ with the boundary conditions $w_i$ and $w_j$, which can be
easily computed by the forward-backward algorithm.
Then we can solve our problem by applying, e.g., Dijkstra's algorithm to
this graph.
Dijkstra's complexity is  bounded from above by the square of the number of nodes and is, thus,
polynomial with respect to the size and the values of data of the original problem,
which proves pseudo-polynomiality.\\

\bibliographystyle{authordate1}
\bibliography{biblio}

\end{document}